\documentclass[letterpaper, 10 pt, conference]{ieeeconf}  

\IEEEoverridecommandlockouts                              
\usepackage{syntax}
\usepackage{url}
\usepackage{graphicx}
\usepackage{amssymb}

\usepackage{amsfonts}
\usepackage[english]{babel}
\usepackage[bottom]{footmisc}
\usepackage{algorithm}
\usepackage{algpseudocode}
\usepackage{amsthm}
\usepackage{siunitx}
\usepackage{array}
\usepackage{verbatim}
\usepackage{subcaption}
\usepackage{listings}
\usepackage{amsmath}
\usepackage{adjustbox}
\usepackage{csquotes}
\usepackage{microtype}
\usepackage{graphicx}
\usepackage{amsmath}
\usepackage{todonotes}
\usepackage{xcolor}
\usepackage{cite}
\usepackage{amsfonts}
\usepackage{xspace} 
\usepackage{multicol}
\graphicspath{{images/}}

\newtheorem{theorem}{Theorem}

\newtheorem{problem}{Problem}

\newtheorem{assumption}{Assumption}
\newtheorem{proposition}{Proposition}

\newtheorem{lemma}{Lemma}
\newtheorem{definition}{Definition}
\newtheorem{remark}{Remark}

\title{\LARGE \bf
Safety Verification of Unknown Dynamical Systems via\\ Gaussian Process Regression
}

\author{John Jackson$^1$, Luca Laurenti$^2$, Eric Frew$^1$, and Morteza Lahijanian$^1$
\thanks{This work was supported in part by the University of Colorado Boulder Autonomous Systems Interdisciplinary Research Theme, and the NSF Center for Unmanned Aircraft Systems, and the National Science Foundation under award IIP - 1650468.}
\thanks{$^1$J. Jackson, E. Frew, and M. Lahijanian are with the Department of Smead Aerospace Engineering Sciences,
        University of Colorado Boulder, CO, USA.
        {\tt \small \{john.m.jackson, eric.frew, morteza.lahijanian\}@colorado.edu}
        }%
\thanks{$^2$L. Laurenti is with the Department of Computer Science, University of Oxford, UK.
        {\tt \small luca.laurenti@cs.ox.ac.uk}}%
}

\renewcommand{\phi}{\varphi}

\newcommand{\set}[1]{\{#1\}}
\newcommand{\reals}{\mathbb{R}}
\newcommand{\naturals}{\mathbb{N}}

\newcommand{\kernel}{\kappa}

\newcommand{\mdp}{\textsc{mdp}\xspace}
\newcommand{\imdp}{\textsc{imdp}\xspace}
\newcommand{\M}{\mathcal{M}}
\newcommand{\I}{\mathcal{I}}

\newcommand{\probDist}{\mathcal{D}}
\newcommand{\FeasibleDist}[2]{\Gamma_{#1}^{#2}}
\newcommand{\feasibleDist}[2]{\gamma_{#1}^{#2}}

\newcommand{\Amdp}{A}
\newcommand{\Qmdp}{Q}
\newcommand{\Pmdp}{P}
\newcommand{\amdp}{a}
\newcommand{\qmdp}{q}

\newcommand{\Qimdp}{Q}
\newcommand{\Aimdp}{A}
\newcommand{\qimdp}{q}
\newcommand{\qimdpprime}{\qimdp^\prime}
\newcommand{\aimdp}{a}

\newcommand{\Pup}{\hat{P}}
\newcommand{\Plow}{\check{P}}

\newcommand{\pathmdp}{\omega}
\newcommand{\pathimdp}{\omega}
\newcommand{\pathmdpfin}{\pathmdp^{\mathrm{fin}}}
\newcommand{\pathimdpfin}{\pathimdp^{\mathrm{fin}}}
\newcommand{\Pathmdp}{\mathit{Paths}}
\newcommand{\Pathmdpfin}{\mathit{Paths}^{\mathrm{fin}}}

\newcommand{\Pathimdpfin}{\mathit{Paths}^{\mathrm{fin}}}
\newcommand{\last}{\mathit{last}}

\newcommand{\str}{\pi}

\newcommand{\adv}{\gamma}

\newcommand{\strX}{\str_{\mathbf{x}}}

\newcommand{\safe}{\mathrm{safe}}
\newcommand{\unsafe}{\mathrm{u}}

\newcommand{\p}{p}
\newcommand{\plow}{\check{\p}}
\newcommand{\pup}{\hat{\p}}

\newcommand{\D}{\mathcal{D}}

\newcommand{\xb}{\textnormal{\textbf{x}}}   
\newcommand{\yb}{\textnormal{\textbf{y}}}
\newcommand{\ub}{\textnormal{\textbf{u}}}

\newcommand{\x}{\xb}
\renewcommand{\u}{\ub}
\newcommand{\y}{\yb}
\newcommand{\SafeSet}{\mathcal{X}_{\safe}}

\newcommand{\distBound}{\epsilon}

\newcommand{\noise}{\mathbf{v}}
\newcommand{\dataset}{\mathrm{D}}
\newcommand{\distribution}{\D}

\newcommand{\compactSet}{\mathcal{K}}

\begin{document}

\maketitle
\thispagestyle{empty}
\pagestyle{empty}

\begin{abstract}
The deployment of autonomous systems that operate in unstructured environments necessitates algorithms to verify their safety. This can be challenging due to, e.g., black-box components in the control software, or undermodelled dynamics that prevent model-based verification. We present a novel verification framework for an unknown dynamical system from a given set of noisy observations of the dynamics. Using Gaussian processes trained on this data set, the framework abstracts the system as an uncertain Markov process with discrete states defined over the safe set. The transition bounds of the abstraction are derived from the probabilistic error bounds between the regression and underlying system. An existing approach for verifying safety properties over uncertain Markov processes then generates safety guarantees. We demonstrate the versatility of the framework on several examples, including switched and nonlinear systems.
\end{abstract}

\section{INTRODUCTION}
    \label{sec:intro}

The ability to provide formal guarantees is essential for \textit{safety-critical} systems.  Without assurances, innovations such as self-driving cars, medical robotics, and autonomous aerial vehicles will remain bounded to narrow domains.  To address this need, formal verification offers powerful frameworks with rigorous analysis techniques~\cite{Clarke99,BaierBook2008}. They provide formal guarantees with respect to the system model. In many applications, however, an accurate model of an autonomous system is either unavailable due to, e.g., the use of a black-box controllers, or if available, it is not in a closed form that can be used for formal verification.  This work focuses on this challenge and aims to develop a verification method that can provide safety guarantees for systems with unknown dynamics.

Formal verification of control systems has been widely studied, e.g., \cite{tabuada2009verification,Belta:Book:2017,doyen2018verification,kushner2013numerical,soudjani2015fau,Lahijanian:TAC:2015,laurenti2020formal}. These methods are typically based on model checking algorithms \cite{Clarke99,BaierBook2008}, which take a simple discrete, finite model and return a \textit{yes} or \textit{no} as to whether the model satisfies a given specification.  To bridge the gap between continuous and discrete domains, those works construct an \textit{abstraction}, a finite representation of the control system with a simulation relation \cite{Girard:ITAC:2007}.  This abstraction is in the form of a finite graph if the underlying system is deterministic or a finite Markov process if the underlying system is stochastic. Even though they admit strong guarantees, these methods are model-based and require full knowledge of the system model.  Hence, they cannot be employed for analysis of systems with unknown dynamics.



In the controls literature, a recent body of work is emerging that focuses on data-driven analysis of dynamical systems, e.g., \cite{Dutta:IFAC:2018,Haesaert:Automatica:2017,Kenanian:Automatica:2019,Ahmadi:CDC:2017}.  Those studies assume partial knowledge about the system and provide some performance assurances.  The work in \cite{Haesaert:Automatica:2017} uses techniques based on Bayesian inference
to compute the confidence over a property of interest for partly unknown linear systems.  Work \cite{Kenanian:Automatica:2019} introduces an algorithm based on chance-constrained optimization to provide probabilistic stability guarantees for an unknown switched linear system from a finite number of observations of trajectories.  
Despite their strengths, those data-driven methods assume the unknown model is linear.
Work \cite{Ahmadi:CDC:2017} relaxes this assumption and considers safety assessment of a dynamical system whose model is fully unknown.  The proposed method is based on approximation of the dynamics using a piecewise-polynomial function and safety assessment through barrier certificates.
This safety analysis is sound with respect to the polynomial function but cannot be extended to the underlying system in a straightforward manner. 


A powerful approach to approximate an unknown function is \textit{Gaussian process} (GP) regression \cite{rasmussen2003gaussian}.
GP regression is a Bayesian machine-learning framework, which has been receiving special attention in safety-critical applications due to its ability to capture the uncertainty in the learning process \cite{cardelli2019robustness,berkenkamp2015safe}.
Recent works \cite{srinivas2012information,Germain2016pac,chowdhury2017kernelized, lederer2019uniform}
successfully derive theoretical bounds on the distance between the regressed GP and the underlying (unknown) system.
These results have led to the increased use of GPs in safe learning frameworks, e.g.,  \cite{akametalu2014reachability,sui2015safe,berkenkamp2017safe,polymenakos2019safety}.  
In \cite{akametalu2014reachability,sui2015safe}, the proposed algorithms learn the unknown dynamics as a GP model, which is then used within a reinforcement learning algorithm to learn a reachability policy under safety constraints.  Similarly, \cite{berkenkamp2017safe} introduces a method of learning a policy safely based on GP modeling with stability guarantees.
Nevertheless, it is unclear whether those algorithms designed for learning policies can be employed for formal verification purposes.  

In this work, we focus on the safety verification of control systems with unknown dynamics via GP regression.  
We introduce an algorithm that, given a set of noisy data, generates formal probabilistic guarantees for the unknown system to remain in a given safe set for every initial state.  The algorithm uses a discretization of the safe set and GP regression to construct a finite abstraction with probabilistic bounds.  This abstraction is in the form of an uncertain Markov model that captures all possible behaviors of the unknown system through a derivation for the error bounds between the regression and underlying system.
Then, the algorithm determines the safety probability bounds for the unknown system by performing safety verification on the abstraction.

The main contribution of this work is a framework for formal verification of unknown dynamical systems.  This is the \textit{first} abstraction-based verification technique that does not assume known dynamics to the best of our knowledge. This work lays the theoretical foundation for formal reasoning about unknown systems against complex specifications given, e.g., as \textit{temporal logic} formulas \cite{BaierBook2008}.  Another contribution of the paper is a derivation of probability bounds on the transition from a point to a region for the unknown dynamics.  These bounds are general and hence can be applied to systems with various levels of knowledge about their dynamics.  
Furthermore, we provide a series of case studies to illustrate the power of the method on linear, switched, and nonlinear systems. 

\section{PROBLEM FORMULATION}
    \label{sec:problem}
Consider a controlled dynamical system with noisy observation (measurement) in the form of
\begin{equation}
    \label{Eqn::Process}
    \begin{split}
        & \x(k+1) = f( \x(k),\u(k)),\\
        & \y(k) = \x(k)+ \noise(k,\u(k-1)),\\
    \end{split}
\end{equation}
where
\begin{equation*}
    \x(k) \in  \mathbb{R}^n, \: \u(k) \in \mathcal{U}, \: \y(k) \in \mathbb{R}^n, \: \noise(k,\u(k-1)) \sim \distribution_{\u(k-1)},
\end{equation*}
$f: \reals^n \times U \to \reals^n$ is a possibly non-linear and unknown function that represents the dynamics of the system, $\mathcal{U} = \set{a_1,\ldots,a_{\mathcal{|U|}}}$ is a finite set of actions or control laws, and for each $a \in \mathcal{U},$ $\noise(k,a)$ is a noise term sampled from distribution $\distribution_a$. We assume the noise $\noise$ is an arbitrary zero-mean martingale difference sequence, i.e., for each $k>0$ and $k'<k$ 
$$\mathbb{E}\big[\noise(k,\u(k)) \mid \noise(k',\u(k'))  \big] = 0.$$
We further assume that $\|\noise\| < \sigma$ almost surely for some $\sigma>0$ at each step $k$ and that the noise on the various components of $\x$ is independent, i.e., component $\noise_i$ is independent of $\noise_j$ for $i,j\in \{1,...,n \}$. 

Taking $f$ as completely unknown may lead to an ill-posed problem. 
We employ the following standard assumption \cite{srinivas2012information}, which guarantees $f$ is a well-behaved function that can be approximated using GP regression.

\begin{assumption}
    \label{assump:smoothnes}
    For a compact set $\compactSet\subset \mathbb{R}^n$, let $\kernel:\mathbb{R}^n\times \mathbb{R}^n\to \mathbb{R}_{> 0}$ be a given kernel and $\mathcal{H}_\kernel(\compactSet)$ the reproducing kernel Hilbert space (RKHS) of functions over $\compactSet$ corresponding to $\kernel$ with norm $\| \cdot \|_\kernel$ \cite{srinivas2012information}.  Then, for each $a \in \mathcal{U}$  and $i\in \{1,...,n \},$ $f_i(\cdot,a) \in \mathcal{H}_\kernel(\compactSet)$ and for a constant $B_i>0,$ $\| f_i(\cdot,a) \|_\kernel \leq B_i$, where $f_i$ is the $i$-th component of $f$.
\end{assumption}
\noindent
Assumption~\ref{assump:smoothnes} is a common assumption in GP regression \cite{srinivas2012information} that 
limits the class of functions that can be considered in Process~\eqref{Eqn::Process}. In fact, the class of functions considered strictly depends on the kernel under consideration.
A universal kernel, such as the widely-used squared exponential kernel, has the property that $\mathcal{H}_\kernel(\compactSet)$ is a set which is dense in $\mathcal{C}(\compactSet)$ -- the set of continuous functions over $\compactSet$.  That is, every continuous function over $\compactSet$ can be approximated arbitrarily well by members of $\mathcal{H}_\kernel(\compactSet)$ \cite{steinwart2001influence}. 

Let $\omega_{\x}(k)= x_0 \xrightarrow{u_0} x_1 \xrightarrow{u_1}  \ldots \xrightarrow{u_{k-1}} x_k$, where $x_1,\ldots,x_k \in \reals^n$, be a trajectory of Process \eqref{Eqn::Process} up to time $k$ with the observation (measurement) trajectory $\omega_{\y}(k)=y_0y_1 \ldots y_k$. Then, a \textit{control strategy} $\strX$ is a measurable function that selects an action (control law) at time $k$ for the system given the observation trajectory up to that time, i.e., $\strX(\omega_{\y}(k)) \in  \mathcal{U}$. Note that $\noise$ is a stochastic process. As a consequence,  $\strX$ and $\x$ are stochastic processes. 


\subsection{Problem}
The focus of this paper is the safety analysis of Process~\eqref{Eqn::Process} from a set of samples, each in the form of $(x,u,y)$, where $y$ is an observation of Process~\eqref{Eqn::Process} with state $x$ and input $u$.  
Note that this analysis needs to be probabilistic due to the reasons stated above and the partial knowledge (finite noisy samples) of Process~\eqref{Eqn::Process}.
The focus is specifically on the verification problem, where the goal is to 
check if a given safety probability threshold is guaranteed.
Therefore, the problem is centered on computing the probability range that $\x(k)$ remains safe for a given (possibly unbounded) time horizon under all possible strategies.
This problem is formally defined below.
 \begin{problem}
 \label{ProbForm}
  Let $\dataset=\set{(x_i,u_i,y_i) \mid i \in \set{1,...,n_{\dataset}}}$ be a set of $n_{\dataset}$ samples of Process \eqref{Eqn::Process}.  Then, for a compact safe set $\SafeSet \subset \reals^n$, a time-horizon $T \in \naturals \cup\{\infty\}$, and every $x \in \SafeSet$, compute the bounds of safety probability $P_\safe(x)$ defined by
 \begin{align*}
    p_{\min}(x) = \min_{\strX} \Pr(\forall k \in [0,T], &\; \omega_{\x}(k)\in \SafeSet  \mid \\   
    & \quad \quad \quad \quad \x(0)=x, \strX,\dataset),\\
    p_{\max}(x) = \max_{\strX} \Pr(\forall k \in [0,T], &\; \omega_{\x}(k)\in \SafeSet \mid \\   
    & \quad \quad \quad \quad \x(0)=x, \strX,\dataset).
 \end{align*}
That is, $P_\safe(x) \in [\,p_{\min}(x),p_{\max}(x)]$ for all possible strategies.
\end{problem}


Note that Problem \ref{ProbForm} is not concerned with finding the strategy $\strX$ that maximizes (or minimizes) the safety probability. Rather, it is focused on checking if a given safety probability threshold is guaranteed for all possible strategies.


\subsection{Approach}
Our approach to Problem \ref{ProbForm} is through a discrete abstraction of Process \eqref{Eqn::Process} in a form of an uncertain Markov decision process. A crucial part of the construction of this abstraction is the derivation of the uncertainty bounds for the transition probability of $\x(k)$ to region $\qimdp' \subset \mathbb{R}^n$ given that $\mathbf{x}(k-1)\in \qimdp \subset \mathbb{R}^n$. Section \ref{sec:abstraction} shows how these bounds can be computed by incorporating the uncertainty from the GP learning process.  Intuitively, the regressed GP may not accurately approximate the posterior of Process \eqref{Eqn::Process} since the observation noise $\noise$ is not Gaussian, i.e., $\noise$ is bounded and the fact that only a finite amount of data is available.  A correction term that captures this discrepancy is required.  
Section \ref{sec:verification} proves the correctness of the proposed method.  

\section{PRELIMINARIES}
    \label{sec:prelim}

Our approach is based on GP regression and Markov processes, which are formally defined in this section.

\subsection{Gaussian Process Regression}
\label{sec:gp-reg}

\textit{Gaussian Process} (GP) regression is a non-parametric Bayesian machine learning method \cite{rasmussen2003gaussian}. For an unknown function  $\mathrm{f}:\mathbb{R}^n\to \mathbb{R}$, the basic assumption of GP regression is that $\mathrm{f}$ is a sample from a GP with zero mean\footnote{Extensions with non-zero mean are a trivial generalization \cite{rasmussen2003gaussian}} and covariance $\kernel: \mathbb{R}^n \times \mathbb{R}^n \to \mathbb{R}_{>0} $. 
GP regression is often used when only noisy observations of $\mathrm{f}$ are available in the form $\mathrm{y} = \mathrm{f}(\mathrm{x}) + \mathrm{v}$, where $\mathrm{v}$ is assumed to be normally distributed with variance $\sigma^2$. Note that here $\mathrm{y},\mathrm{v} \in \reals$ are different from $y,v \in \reals^n$.


Consider a data set of noisy samples $\dataset=\{(\mathrm{x}_i,\mathrm{y}_i),i\in \{1,\dots,n_{\dataset} \}\}$.
Let $X$ and $Y$ be ordered vectors with all points in $\dataset$ such that $X_i = \mathrm{x}_i$ and $Y_i = \mathrm{y}_i$.  Further, call $K(X,X)$ the matrix with $K_{i,j}(X_i,X_j)=\kernel(\mathrm{x}_i,\mathrm{x}_j)$, $K(\mathrm{x},X)$ the vector such that $K_{i}(\mathrm{x},X)=\kernel(\mathrm{x},X_i)$, and $K(X,\mathrm{x})$ defined accordingly. Assuming the noise is i.i.d., the predictive distribution of $\mathrm{f}$ at a test point $\mathrm{x}$ is given by the conditional distribution of $\mathrm{f}$, which is Gaussian and with mean $\mu_\dataset$ and variance $\sigma_\dataset^2$ given by
\begin{align*}
      & \mu_{\dataset}(\mathrm{x}) = K(\mathrm{x},X) \big( K(X,X)+ \lambda I_{n_{\dataset}} \big)^{-1} Y   \\
      & \sigma_{\dataset}^2(\mathrm{x}) = \kernel(\mathrm{x},\mathrm{x})-\\
      &\quad \quad \quad \quad \quad K(\mathrm{x},X)\big( K(X,X)+ \lambda I_{n_{\dataset}} \big)^{-1}K(X,\mathrm{x}),
\end{align*}
where $I_{n_{\dataset}}$ is the identity matrix of size $n_{\dataset} \times n_{\dataset}$ and $\lambda$ is a free parameter (often taken to be $\sigma^2$ when $\mathrm{f}$ is distributed according to the posterior).


In our setting, we do not assume that $\mathrm{f}$ is sampled from a GP and noise $\mathrm{v}$ is not Gaussian, so the assumptions for GP regression are not satisfied. 
Nevertheless, Assumption~\ref{assump:smoothnes} permits using GP regression even in our scenario. 
In particular, the following Lemma provides a bound on the distance between $\mu_{\dataset}$ and $\mathrm{f}$ so long Assumption~\ref{assump:smoothnes} holds. This is an important result for safety verification, where the distance between the regression and the true system needs to be considered.




\begin{lemma}[\hspace{1sp}\cite{chowdhury2017kernelized}, Theorem 2]\label{th:RKHS}
Let $\compactSet$ be a compact set, $\delta\in(0,1)$, $\alpha_\dataset$ the maximum information gain parameter associated with $\kernel$ and data set $\dataset$ training points, and $B>0$ such that $\|\mathrm{f} \|_{\kernel}\leq B$. Assume that $|\mathrm{v}|<\sigma$ almost surely and $\mu_D$ and $\sigma_\dataset$ are found with $\lambda=1+2/n_\dataset$. Define $\beta=(\sigma/\sqrt{\lambda})(B + \sigma\sqrt{2(\alpha_\dataset + 1 + \log{1/\delta})})$. Then, it holds that 
\begin{equation*}
    \Pr\big(\forall x \in \compactSet, |\mu_D(\mathrm{x}) - \mathrm{f}(\mathrm{x})| \leq \beta\sigma_D(\mathrm{x}) \big)\geq 1-\delta.
\end{equation*}
\end{lemma}Lemma~\ref{th:RKHS} computes a probabilistic bound between the regressed GP and the underlying unknown function and takes into account the modelling errors in running GP regression with observation noise with the parameter $\lambda$ and scaling factor $(\sigma/\sqrt{\lambda})$. The constraint on $\|\mathrm{f} \|_{\dataset}$ implies $\mathrm{f}$ is $L_\mathrm{f}$-Lipschitz continuous with $L_\mathrm{f}^2\propto B$~\cite{berkenkamp2017safe}. The information gain term $\alpha_\kappa$ can be upper bounded for certain kernel choices as shown in \cite{srinivas2012information}.




\subsection{Markov Processes}
\label{sec:markov-models}

Our abstraction structure is based on Markov models.

\begin{definition}[\mdp] 
    \label{def:mdp}
    A Markov decision process (\mdp) is a tuple $\M = (\Qmdp,\Amdp,\Pmdp)$, where
    \begin{itemize}
        \item $\Qmdp$ is a finite set of states,
        \item $\Amdp$ is a finite set of actions,
        \item $\Pmdp: \Qmdp \times \Amdp \times \Qmdp \rightarrow [0,1]$ is a transition probability function.
    \end{itemize}
\end{definition}
\noindent
We denote the set of actions available at \mdp state $\qmdp \in \Qmdp$ by $\Amdp(\qmdp)$. 

A path $\pathmdp$ of an \mdp is a sequence of states $\pathmdp = \qmdp_0 \xrightarrow{\amdp_0} \qmdp_1 \xrightarrow{\amdp_1} \qmdp_2 \xrightarrow{\amdp_2}  \ldots$ such that $\amdp_i \in \Amdp(\qmdp_i)$ and $\Pmdp(\qmdp_i, \aimdp_i, \allowbreak{\qimdp_{i+1}}) > 0$ for all $i \in \naturals$. We denote 
the last state of a finite path $\pathmdpfin$ by $\last(\pathmdpfin)$ and the set of all finite and infinite paths by $\Pathmdpfin$ and $\Pathmdp$, respectively.

\begin{definition}[Strategy]
\label{def:strategy}
    A strategy $\str$ of an \mdp model $\M$ is
    a function $\str: \Pathmdpfin \rightarrow \Amdp$ that maps a finite path $\pathmdpfin$ of $\M$ onto an action in $\Amdp$.  
\end{definition}

\noindent Given a strategy $\str$, a probability measure over the set of all paths (under $\str$) $\Pathmdp$ is induced on the resulting Markov chain \cite{BaierBook2008}.

When modeling with MDPs, it might be difficult to determine the exact values of transition probabilities between states, especially if the underlying system is unknown. 
In such cases, we may consider an interval for each value.  
The model that allows the inclusion of these intervals is known as  the \textit{bounded-parameter} \cite{givan2000bounded} or \textit{interval} \mdp (\imdp) \cite{Hahn:QEST:2017}, whose formal definition is as follows.   


\begin{definition}[{\imdp}] \label{def:imdp}
    An interval Markov decision process ({\imdp}) is a tuple $\I = (\Qimdp,\Aimdp,\Plow,\Pup)$, where $\Qimdp$, $\Aimdp$ are as in Def. \ref{def:mdp}, and
    \begin{itemize}
    	\setlength\itemsep{1mm}
        \item $\Plow: \Qimdp \times \Aimdp \times \Qimdp \rightarrow [0,1]$ is a function, where $\Plow(\qimdp,\aimdp,\qimdpprime)$ defines the lower bound of the transition probability from state $\qimdp$ to state $\qimdpprime$ under action $\aimdp \in \Aimdp(\qimdp)$,
        \item $\Pup: \Qimdp \times \Aimdp \times \Qimdp \rightarrow [0,1]$ is a function, where $\Pup(\qimdp,\aimdp,\qimdpprime)$ defines the upper bound of the transition probability from state $\qimdp$ to state $\qimdpprime$ under action $\aimdp \in \Aimdp(\qimdp)$.
    \end{itemize}
\end{definition}
\noindent
For all $\qimdp,\qimdpprime \in \Qimdp$ and $\aimdp \in \Aimdp(\qimdp)$, it holds that $\Plow(\qimdp,\aimdp,\qimdpprime) \leq \Pup(\qimdp,\aimdp,\qimdpprime)$ and
\begin{equation*}
    \sum_{\qimdpprime \in \Qimdp} \Plow(\qimdp,\aimdp,\qimdpprime) \leq 1 \leq \sum_{\qimdpprime \in \Qimdp} \Pup(\qimdp,\aimdp,\qimdpprime).
\end{equation*}
Let $\distribution(\Qimdp)$ denote the set of discrete probability distributions over $\Qimdp$.  Given $\qimdp \in \Qimdp$ and $\aimdp \in \Aimdp(\qimdp)$, we call $\feasibleDist{\qimdp}{\aimdp} \in \probDist(\Qimdp)$ a \textit{feasible distribution} reachable from $\qimdp$ by $\aimdp$ if 
\begin{equation*}
    \Plow(\qimdp,\aimdp,\qimdpprime) \leq \feasibleDist{\qimdp}{\aimdp}(\qimdpprime) \leq \Pup(\qimdp,\aimdp,\qimdpprime) 
\end{equation*}
for each state $\qimdpprime \in \Qimdp$.
We denote the set of all feasible distributions for state $\qimdp$ and action $\aimdp$ by $\FeasibleDist{\qimdp}{\aimdp}$. 

The notions of paths and strategies of {\imdp}s are analogous to those of {\mdp}s.  An additional notion is the \textit{adversary} that chooses feasible distributions.


\begin{definition}[Adversary]
\label{def:adversary}
    Given an \imdp  $\,\I$, an adversary is a function $\adv: \Pathimdpfin \times \Aimdp \rightarrow \probDist(\Qimdp)$ that, for each  finite path $\pathimdpfin \in \Pathimdpfin$ and action $\aimdp \in \Aimdp(\last(\pathimdpfin))$, assigns a feasible distribution $\adv(\pathimdpfin,\aimdp) \in \FeasibleDist{\last(\pathimdpfin)}{\aimdp}$. 
\end{definition}


Given a strategy $\str$ and an adversary $\adv$, a Markov chain is resulted from an \imdp.  This Markov chain defines a probability measure over the paths of the \imdp \cite{Lahijanian:TAC:2015}.

\section{ABSTRACTION}
    \label{sec:abstraction}
\label{sec:FullAbstraction}

In order to solve Problem \ref{ProbForm}, we abstract Process \eqref{Eqn::Process} as an \imdp $\I = (\Qimdp,\Aimdp,\Plow,\Pup)$ as detailed below.  

\subsection{States \& Actions}
\label{sec:imdp-states-actions}

First, we partition the compact safe set $\SafeSet$ into a set of cells (regions) that are non-overlapping.
Let $\Qimdp_\safe =\set{\qimdp_1,...,\qimdp_{|\Qimdp_\safe|}}$ be the resulting set of cells.  
Then, 
$\cup_{\qimdp \in \Qimdp_\safe} \qimdp = \SafeSet,$
and
\begin{equation*}
	\qimdp \cap \qimdp' = \emptyset, \quad \forall \qimdp,\qimdp' \in \Qimdp_\safe, \text{ and } \qimdp \neq \qimdp'.
\end{equation*}
Each region is associated to a state of \imdp $\I$.  With an abuse of notation, $\qimdp$ denotes both the region, i.e., $\qimdp \subset \SafeSet$, as well as its corresponding \imdp state, i.e, $\qimdp \in \Qimdp$.  From the context, the correct interpretation of $\qimdp$ should be clear.  Furthermore, let $q_\unsafe$ denote the unsafe set $\reals^n \setminus \SafeSet$.  Then, the set of states of $\I$ is defined as
\begin{equation*}
	\Qimdp = \Qimdp_\safe \cup \set{\qimdp_\unsafe}.
\end{equation*}

The set of actions $\Aimdp$ of $\I$ is given by the set of actions in $\mathcal{U}$, and all actions are allowed to be available at each state of $\I$, i.e., $\Aimdp(\qimdp) = \Aimdp$ for all $\qimdp \in \Qimdp$.

\subsection{Transition Probability Bounds}
\label{sec:TransitionProbabilities}
In order to account for the probabilistic behavior of Process \eqref{Eqn::Process}, we define the following conditions for the transition probability bounds of $\I$:
\begin{align}
	\label{eqn:CriteriaTransitionLowerBound}
    \Plow(\qimdp,\aimdp,\qimdp') &\leq \min_{x\in \qimdp} \Pr(\x(k) \in \qimdp' \mid \x(k-1)=x, \nonumber \\ 
    & \hspace{35mm}  \u(k-1)=a,\dataset), \\
    \label{eqn:CriteriaTransitionUpperBound}
    \Pup(\qimdp,\aimdp,\qimdp') &\geq \max_{x\in \qimdp} \Pr(\x(k) \in \qimdp' \mid \x(k-1)=x, \nonumber \\ 
    & \hspace{35mm}  \u(k-1)=a,\dataset),
\end{align}
for all $\qimdp,\qimdp' \in Q$. Notice that even though the action is fixed in \eqref{eqn:CriteriaTransitionLowerBound} and \eqref{eqn:CriteriaTransitionUpperBound}, a probabilistic statement is necessary because $f$ is unknown and
the samples in $D$ are noisy. 
Conditions \eqref{eqn:CriteriaTransitionLowerBound} and \eqref{eqn:CriteriaTransitionUpperBound} guarantee that the full probabilistic behavior of Process \eqref{Eqn::Process} is accounted for in the abstraction as shown in Section \ref{sec:verification}.
In order to compute the bounds that satisfy these conditions, we partition the set of samples $\dataset$ according to actions $\aimdp \in \mathcal{U}$, i.e.,  
$\dataset = \cup_{\aimdp \in \mathcal{U}} \dataset_a,$
where
$$\dataset_a = \set{(x_i,a,y_i) \mid (x_i,a,y_i) \in \dataset}.$$
GP regression on $\dataset_a$ for each $a$ results in a Gaussian posterior distribution characterized by mean $\mu_{\dataset}^a$ and diagonal covariance matrix $\Sigma_{\dataset}^{a}$.
Recall that, even though $f$ is unknown and $\noise$ is not Gaussian, Lemma \ref{th:RKHS} allows one to characterize the distance between the posterior mean $\mu_{\dataset}^a$ and $f(\cdot,a).$


\subsubsection{Transitions to Safe States}
For all the safe states $\qimdp,\qimdp' \in \Qimdp_\safe$, the transition probability bounds in \eqref{eqn:CriteriaTransitionLowerBound} and \eqref{eqn:CriteriaTransitionUpperBound} are given by Proposition \ref{Th:Strategy}. In order to state this result, we introduce the notions of reduction and enlargement of a compact set. 

For a scalar $\distBound >0$ and a compact set $q \subset \mathbb{R}^n$, let $\underline{q} \subset q$ be a subset of $q$ such that the distance between each of its points to the boundary of $q$ is at least $\distBound$.  Moreover, let $\overline{q}$ be such that $q \subset \overline{q}$ and $\overline{q}$ contains all the points that are within a $\distBound$ margin from the boundary of $q$.  Sets $\underline{q}$ and $\overline{q}$ are the $\distBound$-\textit{reduced} and $\distBound$-\textit{enlarged} versions of $q$, respectively.  We are now ready to state the following result:
\begin{proposition}
	\label{Th:Strategy}
	Let  $\qimdp, \qimdp' \subset \mathbb{R}^n$ be compact sets. For $\distBound>0$, define the enlarged and reduced sets
	$$\overline{\qimdp}'=\{ x \in \mathbb{R}^n \mid \, \exists x'\in \qimdp' \;\; s.t. \;\; \|x-x'\|_{\infty}\leq \distBound   \}$$ 
	and 
	$$\underline{\qimdp}'=\{ x \in \qimdp' \mid \, \forall x'\in \partial \qimdp', \; \|x-x'\|_{\infty} > \distBound  \}, $$ 
	where $\partial \qimdp'$ is the boundary of $\qimdp'$. Then, for a given action $a \in \mathcal{U}$, it holds that
	\begin{align*}
		& \min_{x\in \qimdp} \Pr(\x(k)\in \qimdp' \mid \x(k-1)=x, \u(k-1)=a,\dataset) \geq \\
		& \quad \min_{x \in \qimdp} \Big( \mathbf{1}_{ \underline{\qimdp}' }(\mu^{a}_\dataset (x))  \\
		&\hspace{13mm}\prod_{i=1}^n \Pr( \forall x'\in \qimdp, |f_i(x',a) - \mu_{i,\dataset}^a(x')| \leq \distBound \mid  \dataset ) \big),
	\end{align*}
	and
	\begin{align*}
		& \max_{x\in \qimdp} \Pr(\x(k)\in \qimdp' \mid \x(k-1)=x, \u(k-1)=a,\dataset) \leq     \\
		& \quad \max_{x \in \qimdp} \Big( 1 -  \prod_{i=1}^n \Pr \big(\forall x'\in \qimdp, \, |f_i(x',a) - \mu_{i,\dataset}^a(x') | \leq \distBound \mid  \\
		& \hspace{50mm} \dataset \big) \big(1- \mathbf{1}_{ \overline{\qimdp}'}(\mu^{a}_\dataset (x)) \big)  \Big),
	\end{align*}
	where $\mu_{i,\dataset}^a$ is the $i$-th component of vector $\mu_{\dataset}^a$, and $\mathbf{1}_H(h)$ is the indicator function which is $1$ if $h\in H$ and $0$ otherwise.
\end{proposition}
\noindent

Proposition~\ref{Th:Strategy} guarantees that upper and lower bounds of $\Pr( \x(k) \in \qimdp' \mid  \x(k-1) \in \qimdp , \u(k-1 ) = a, \dataset )$ can be derived by checking if the posterior mean is within a reduced (or enlarged) version of $\qimdp'$ and computing a uniform error bound on the distance between the posterior mean of the GP learnt via GP regression and $f$ (the underlying dynamics in Process \eqref{Eqn::Process}).
Such a bound can be computed by employing Lemma \ref{th:RKHS}. Proposition \ref{Th:Strategy} and Lemma \ref{th:RKHS} are combined in the following theorem.
\begin{theorem}\label{Th;FInalwithRKHS}
Let $\qimdp,\qimdp',\overline{\qimdp}',\underline{\qimdp}'$ be as defined in Proposition \ref{Th:Strategy}. For $i\in \{1,...,n\}$ consider $B_i>0$ such that for a given $a\in \mathcal{U},$ $\|f(\cdot,a)\|_{\kernel}\leq B_i.$ Define $\beta_i$ as in Lemma~\ref{th:RKHS} and for $\delta\in(0,1)$ select $\epsilon=\max_{i\in \{1,...,n\}}\beta_i^{\frac{1}{2}}(\Sigma_{\dataset}^{a,(i,i)})^{\frac{1}{2}},$ where $\Sigma_{\dataset}^{a,(i,i)}$ is the $i$-th element of the diagonal of $\Sigma_{\dataset}^{a}$. Then, it holds that
	\begin{multline*}
		\min_{x\in \qimdp} \Pr(\x(k)\in \qimdp' \mid \x(k-1)=x, \u(k-1)=a,\dataset) \geq \\
		\min_{x \in \qimdp}  \mathbf{1}_{ \underline{\qimdp}' }(\mu^{a}_\dataset (x))(1-\delta)^n,
	\end{multline*}
	and
	\begin{multline*}
		\max_{x\in \qimdp} \Pr(\x(k)\in \qimdp' \mid \x(k-1)=x, \u(k-1)=a,\dataset) \leq     \\
		\max_{x \in \qimdp} \Big( 1 -  (1-\delta)^n(1- \mathbf{1}_{ \overline{\qimdp}'}(\mu^{a}_\dataset (x)) )  \Big).
    \end{multline*}
\end{theorem}
\noindent
The proof is obtained by directly applying Lemma \ref{th:RKHS} to Proposition \ref{Th:Strategy}.


Note that Theorem \ref{Th;FInalwithRKHS} holds for every choice of constant $\delta$, and hence of $\epsilon$. As discussed in Section \ref{sec:evaluation}, this constant should be selected in order to maximize the tightness of the bound.
In fact, $\epsilon$ controls both the tightness of the bound between the posterior mean and the underlying system and how much $\qimdp'$ is reduced and enlarged. 
\begin{remark}
\label{remark:GeneralityBounds}
Proposition \ref{Th:Strategy} is general; in that, it does not make use of Assumption \ref{assump:smoothnes}. It just assumes the existence of a bound between the posterior mean and the unknown function.  
Therefore, it can be applied to other settings where Assumption \ref{assump:smoothnes} is not satisfied.  
For instance, if function $f$ is a sample from a GP prior with Gaussian observation noise, our framework can still be used in combination with existing error bounds developed for this scenario, such as those in \cite{lederer2019uniform}.
\end{remark}

\subsubsection{Transitions to Unsafe State}
We obtain upper and lower bounds for the transitions to the unsafe region $q_\unsafe$ as a corollary of Theorem \ref{Th:Strategy}. That is, for every $q \in Q_\safe$,
\begin{align*}
    \Plow(\qimdp,\aimdp,&q_u) = 1- \max_{x \in q} \Pr(\x(k) \in \SafeSet \mid \\
    & \hspace{27mm} \x(k-1)=x, \u(k-1)=a, \dataset),\\
    \Pup(\qimdp,\aimdp,&q_u) = 1- \min_{x \in q} \Pr(\x(k) \in \SafeSet \mid  \\
    & \hspace{27mm} \x(k-1)=x, \u(k-1)=a, \dataset).
\end{align*}
Both of these terms can be computed by employing Theorem \ref{Th;FInalwithRKHS}. To complete the construction of abstraction $\I$, we make $q_\unsafe$ absorbing, i.e., $\Plow(q_\unsafe,a,q_\unsafe)= \Pup(q_\unsafe,a,q_\unsafe) = 1$ for all $a \in A$, to ensure that  $\I$ does not count the transitions to $\SafeSet$ from $q_\unsafe$ of Process \eqref{Eqn::Process} as a safe behavior.  

\section{VERIFICATION}
    \label{sec:verification}

Given the \imdp abstraction $\I$, we are interested in computing the probabilities of remaining in $\Qimdp_\safe$ for $T \in \naturals \cup \set{\infty}$ time steps from every $q \in Q_\safe$.  Note that, under strategy $\str$, the safety probability is a range due to the transition probability intervals of $\I$.  The values in this range correspond to the feasible transition probabilities $\feasibleDist{q}{a} \in \FeasibleDist{q}{a}$ at every state $q\in Q$ and action $a \in A$ chosen by $\str$.  The choice of this feasible transition probability is made by adversary $\adv$.  Therefore, the minimum safety probability is achieved when both strategy $\str$ and adversary $\adv$ are minimizing the safety probability.  Similarly, the maximum safety probability is given when both $\str$ and $\adv$ are maximizing.  

This optimization problem can be formulated through the Bellman equation as detailed in \cite{Lahijanian:TAC:2015}.  Let $\plow^k(q)$ and $\pup^k(q)$ denote the minimum and maximum probability of remaining safe in $k$ time steps starting from state $\qimdp \in \Qimdp$, respectively.  Then, the safety probability bounds for a finite time duration $T$ can be computed by $T$ recursive evaluations of
\begin{align}
	\label{eq:safety-bellman-lower}
	\plow^k(\qimdp) = \min_{\aimdp \in \Aimdp(\qimdp)} \; \min_{\feasibleDist{\qimdp}{\aimdp} \in \FeasibleDist{\qimdp}{\aimdp} } \; \sum_{\qimdp' \in \Qimdp} \feasibleDist{\qimdp}{\aimdp}(\qimdp') \; \plow^{k-1}(\qimdp') \\
	\label{eq:safety-bellman-upper}
	\pup^k(\qimdp) = \max_{\aimdp \in \Aimdp(\qimdp)} \; \max_{\feasibleDist{\qimdp}{\aimdp} \in \FeasibleDist{\qimdp}{\aimdp} } \; \sum_{\qimdp' \in \Qimdp} \feasibleDist{\qimdp}{\aimdp}(\qimdp') \; \plow^{k-1}(\qimdp')
\end{align}
with initial values of $\plow^0(\qimdp) = 1$ for $\qimdp \in \Qimdp_\safe$ and $\plow^0(\qimdp_\unsafe) = 0$.  In the case of an infinite time horizon $T$, recursive evaluations of \eqref{eq:safety-bellman-lower} and \eqref{eq:safety-bellman-upper} need to continue until convergence, which is guaranteed to occur in finite time \cite{Lahijanian:TAC:2015}.

This method of evaluation is similar to value iteration.  The additional step involves first optimizing over the adversaries, which itself can be performed iteratively via an ordering of the states in $\Qimdp$ according to their values \cite{Lahijanian:TAC:2015}.  
Once the optimal adversaries are obtained for all $\aimdp \in \Aimdp(q)$,  an optimization over the actions is performed to complete the computation for one time step in \eqref{eq:safety-bellman-lower} and \eqref{eq:safety-bellman-upper}.
This algorithm computes the safety probability bounds $\plow^{T}(\qimdp)$ and $\pup^{T}(\qimdp)$ for each $\qimdp \in \Qimdp$.  The complexity of the algorithm is polynomial in the size of the \imdp $\I$ \cite{Lahijanian:TAC:2015}.


\subsection{Correctness}
The following theorem guarantees that the safety probability ranges computed by our framework are sound, i.e., they give lower and upper bounds for $p_{\min}(x)$ and $p_{\max}(x)$ as defined in Problem \ref{ProbForm}.
\begin{theorem}
    \label{th:correctness}
    Let $ x \in \SafeSet$ and $\qimdp\in \Qimdp$ such that $x\in \qimdp$. Then, it holds that
    $$ [p_{\min}(x),p_{\max}(x)]\subseteq \big[\plow^T(\qimdp), \, \pup^T(\qimdp) \big].$$
\end{theorem}
\noindent

\section{CASE STUDIES}
    \label{sec:evaluation}
    
We evaluate the performance of our framework in three case studies. The first case study involves three single-action linear systems and shows the effect of various choices for parameter $\distBound$. The second case uses two of the linear systems to define a switched system with two actions. The final case considers the safety a nonlinear system. 


In all three case studies, $\SafeSet$ is a two-dimensional square defined by the region $\SafeSet=[-4, 4]\times[-4, 4]$. 
We performed a regression of each dynamical system using a pair of Gaussian processes, one for each output component. The GP prior used the zero mean and squared-exponential functions. The training process used one thousand i.i.d. training points with noise parameter $\sigma=0.01$ to optimize the hyperparameters of the prior functions and train the Gaussian processes using the GaussianProcesses.jl Julia package~\cite{gaussianprocesses.jl}. We modified an existing tool to perform the verification over the resulting \imdp \cite{Lahijanian:TAC:2015}.


\begin{figure}
    \centering
    \includegraphics[width=0.23\textwidth]{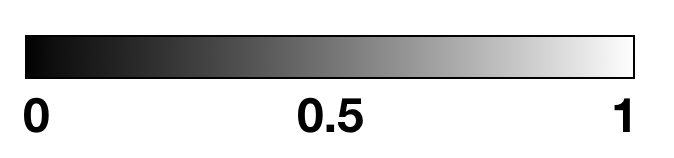}\\
    \subcaptionbox{$p_{\min}$ for $A_\text{rotation}$}[0.23\textwidth][c]{%
    \includegraphics[width=0.23\textwidth]{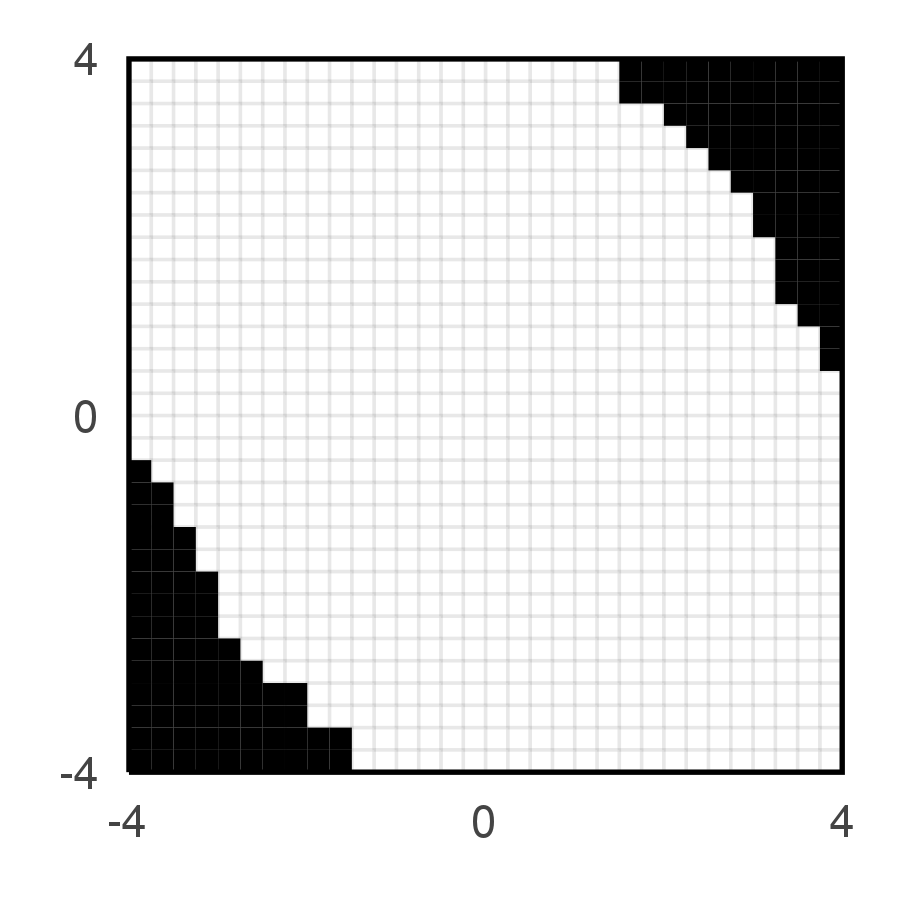}}
    \subcaptionbox{$p_{\max}$ for $A_\text{rotation}$}[0.23\textwidth][c]{%
    \includegraphics[width=0.23\textwidth]{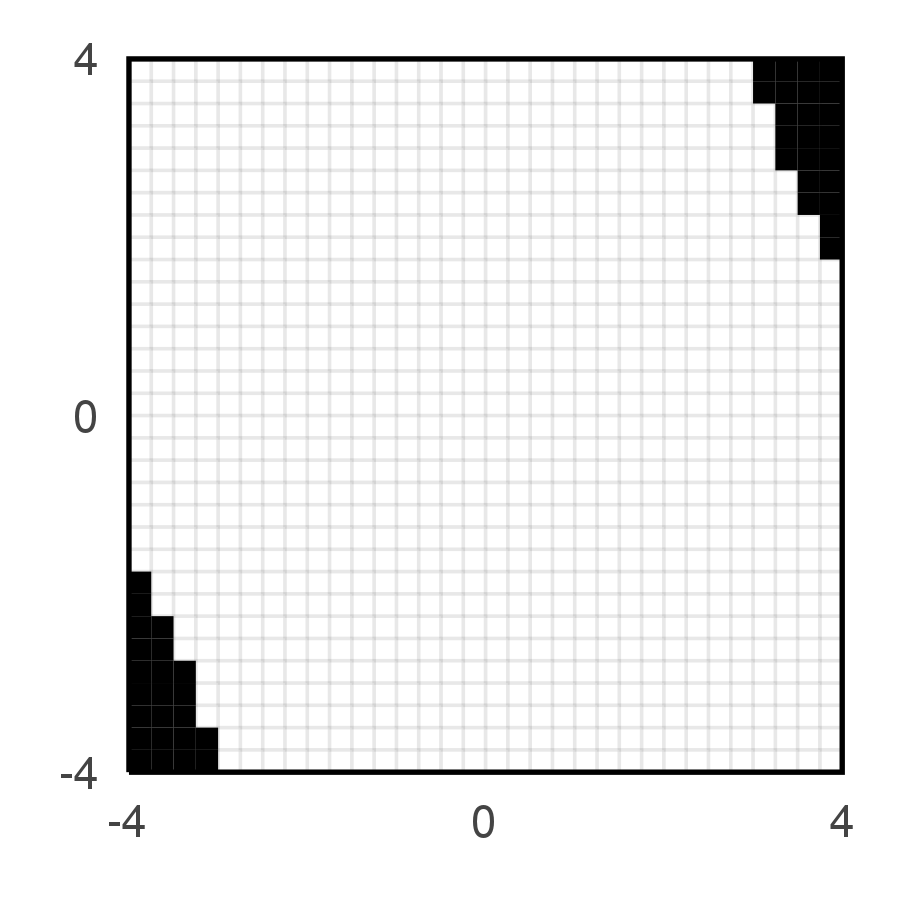}}\\
    \subcaptionbox{$p_{\min}$ for $A_\text{upper}$}[0.23\textwidth][c]{%
    \includegraphics[width=0.23\textwidth]{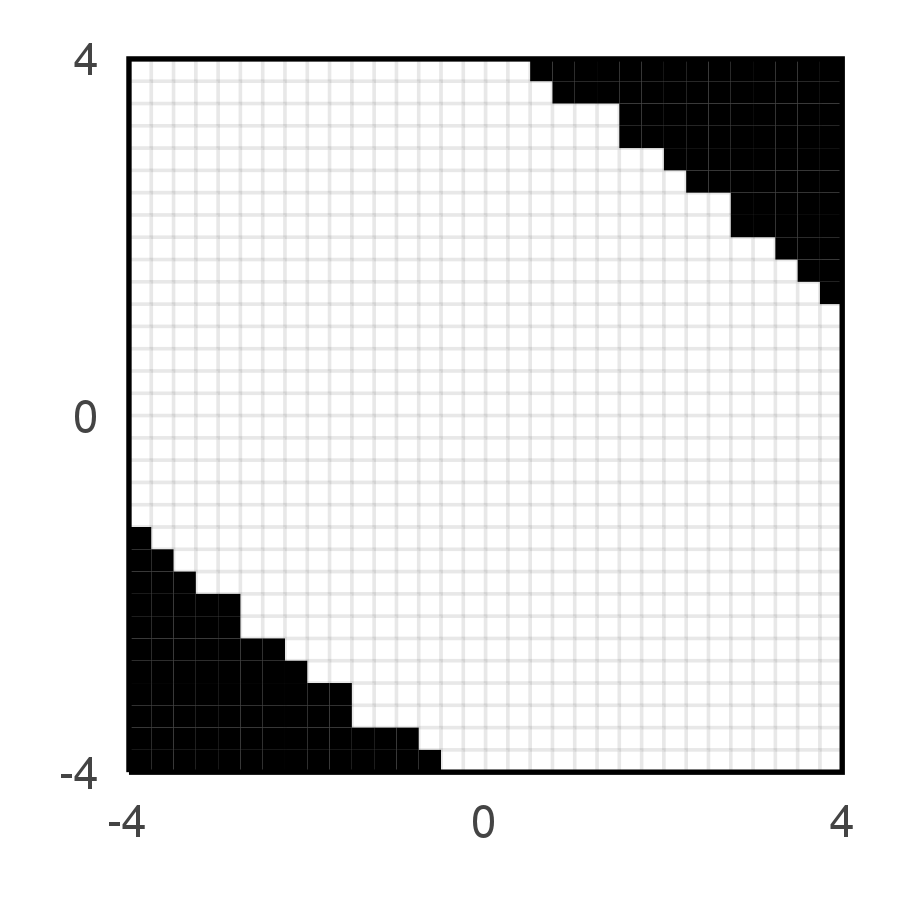}}
    \subcaptionbox{$p_{\max}$ for $A_\text{upper}$}[0.23\textwidth][c]{%
    \includegraphics[width=0.23\textwidth]{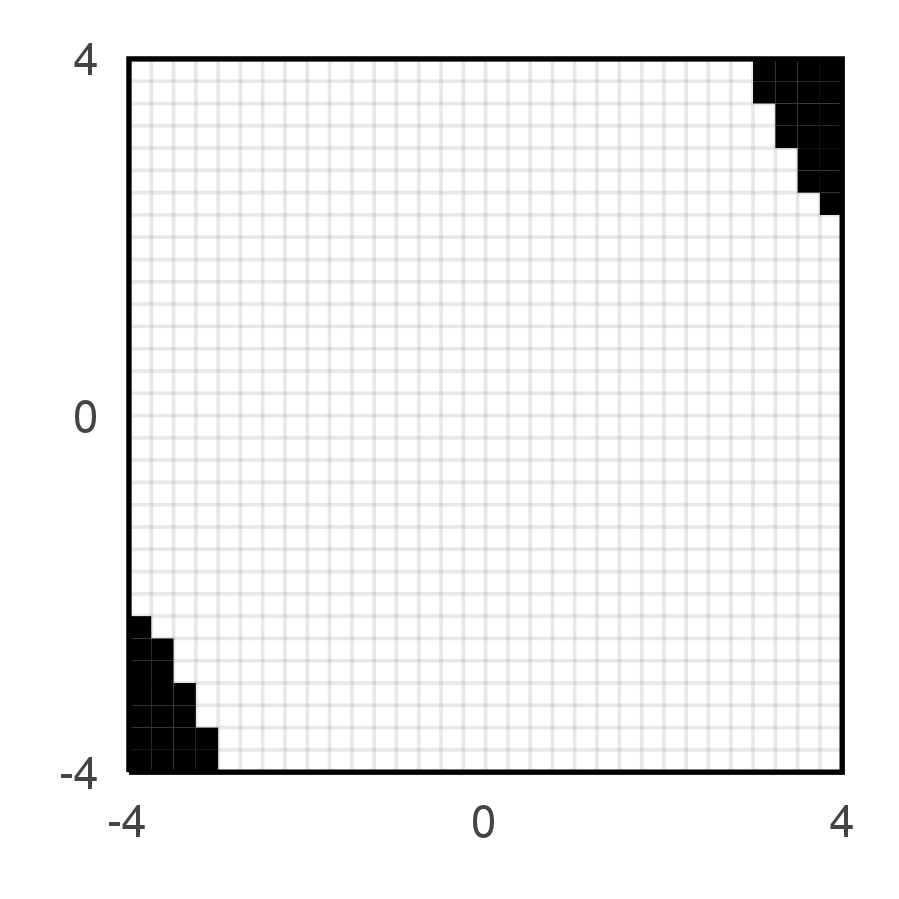}}\\
    \subcaptionbox{$p_{\min}$ for $A_\text{lower}$}[0.23\textwidth][c]{%
    \includegraphics[width=0.23\textwidth]{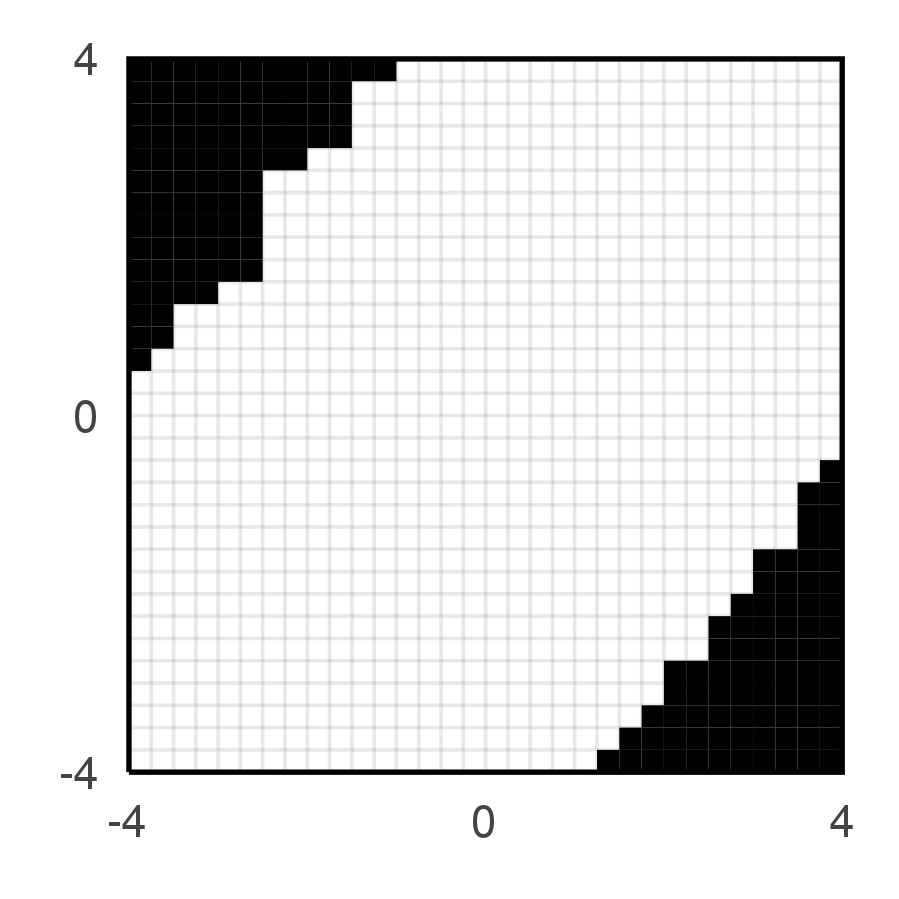}}
    \subcaptionbox{$p_{\max}$ for $A_\text{lower}$}[0.23\textwidth][c]{%
    \includegraphics[width=0.23\textwidth]{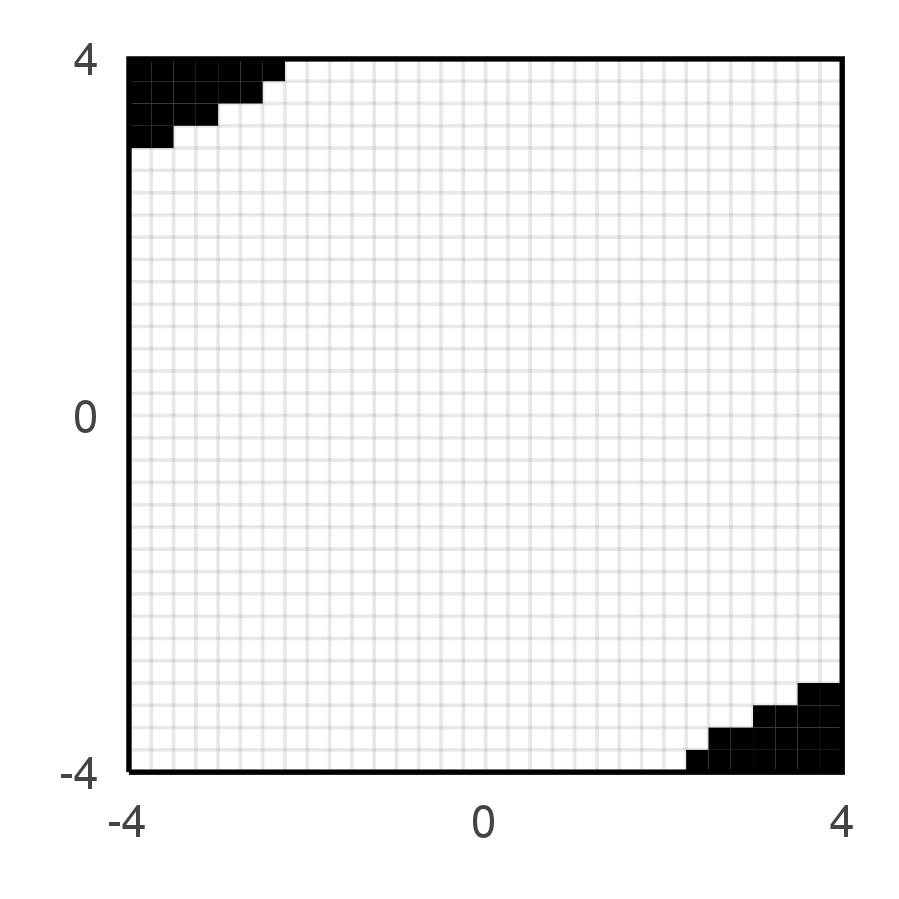}}
\caption{Minimum and maximum 10-step ($T=10$) safety probability for the linear systems with $\distBound=0.12$.}
\label{fig:linear-sys}
\end{figure}

\subsection{Single-Action Linear Systems}
We performed the verification procedure on three linear systems $f(\x(k))=A_i \, \x(k)$, $i \in \set{\text{rotation}, \text{upper}, \text{lower} }$, where
\begin{align*}
    A_{\text{rotation}} = \begin{bmatrix}0.9 & -0.4 \\ 0.4 & 0.5 \end{bmatrix},\qquad\qquad\qquad\\
    A_{\text{upper}} = \begin{bmatrix}0.8 & 0.5 \\ 0 & 0.5 \end{bmatrix},~~~~~~A_{\text{lower}} = \begin{bmatrix}0.5 & 0 \\ -0.5 & 0.8 \end{bmatrix}.
\end{align*}
We discretized the safe set $\SafeSet$ into squares with side length 0.25. Figure~\ref{fig:linear-sys} shows the 10-step safety probability ($T=10$) for each cell using $\distBound=0.12$. The legend above the figures maps the intensity of the shade of each cell to a probability value between zero and one. The white cells in Figures~\ref{fig:linear-sys}(a), (c) and (e) correspond to a minimum safety probability of one. If the system is initialized within one of these cells, then it is certain to remain in the safe set.


Figures~\ref{fig:linear-sys}(a), (c) and (e) also include cells where the minimum probability of safety is zero due to flow that leaves the $\SafeSet$ before returning. This does not necessarily imply that it is impossible to stay in the safe set starting at one of these cells, because the maximum probability of safety may be greater than zero. The maximum probability for three systems is shown in Figure~\ref{fig:linear-sys}(b), (d) and (f). These results indicate that if the system were initialized in the cells with a maximum probability of safety near zero (e.g. the black corners in (b), (d) and (f)), it is certain to leave $\SafeSet$. Cells with a safety probability minimum of zero and maximum of one essentially indicate a nondeterministic transition to a safe or unsafe cell after $T$ steps.


\begin{figure}
    \centering
    \includegraphics[width=0.23\textwidth]{images/scale.png}\\
    \subcaptionbox{$\distBound=0.08$}[0.23\textwidth][c]{%
    \includegraphics[width=0.23\textwidth]{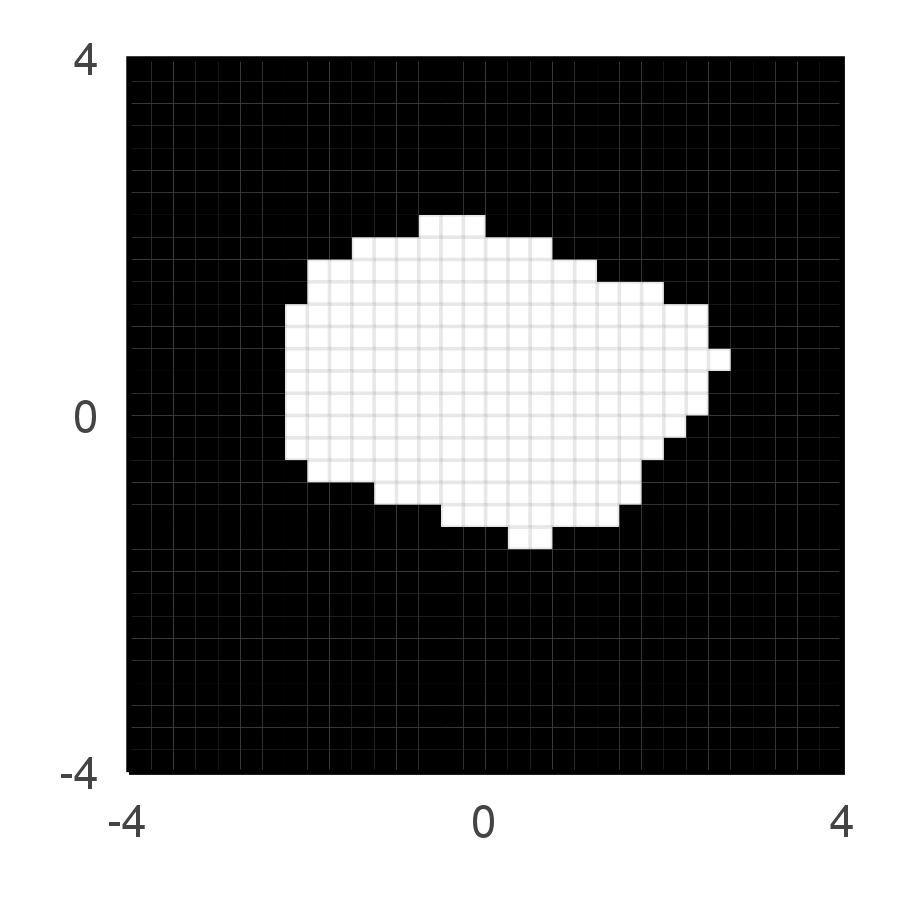}}
    \subcaptionbox{$\distBound=0.09$}[0.23\textwidth][c]{%
    \includegraphics[width=0.23\textwidth]{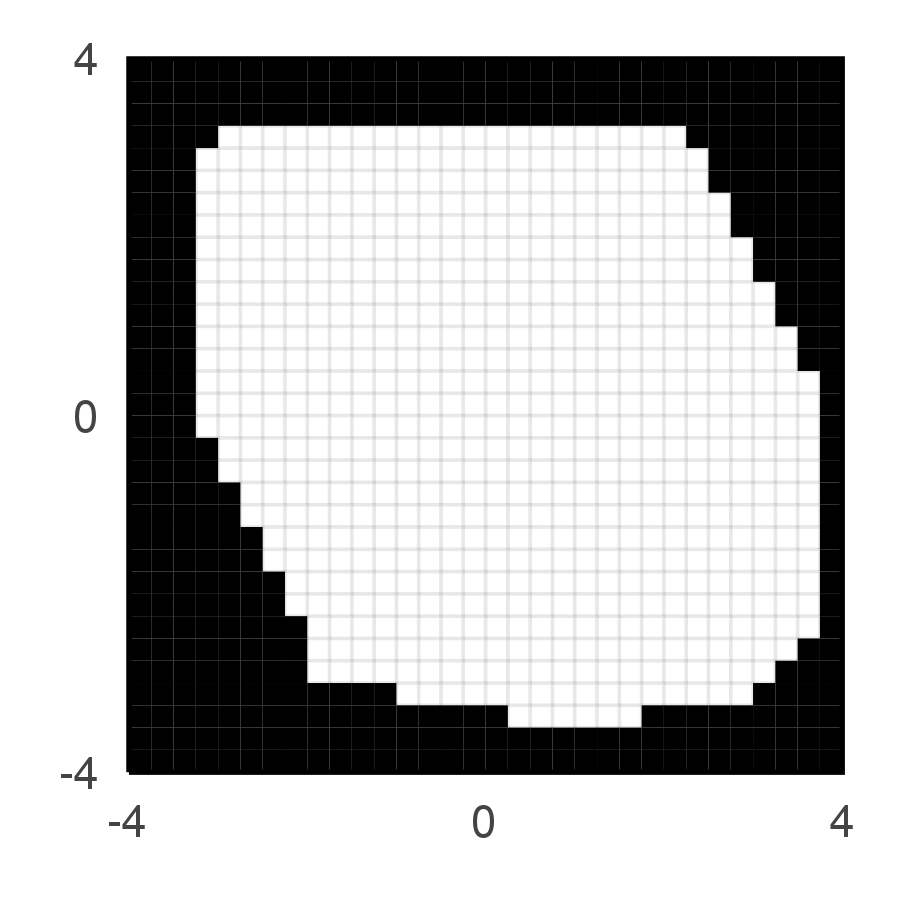}}\\
    \subcaptionbox{$\distBound=0.10$}[0.23\textwidth][c]{%
    \includegraphics[width=0.23\textwidth]{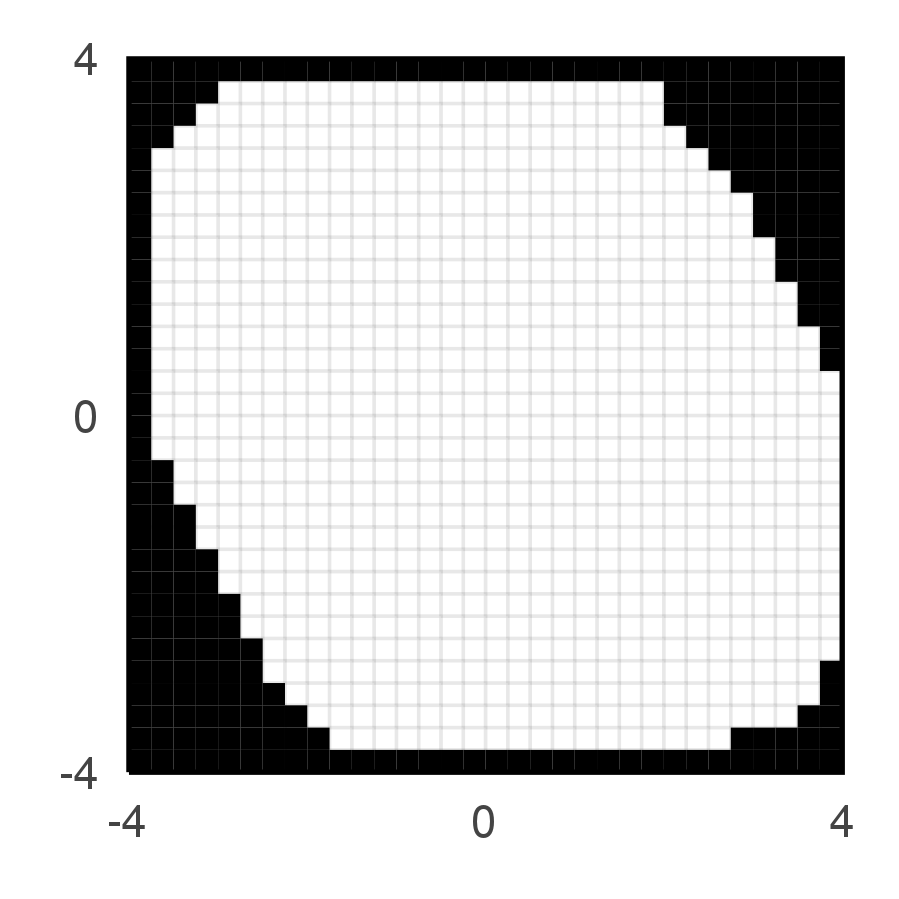}}
    \subcaptionbox{$\distBound=0.12$}[0.23\textwidth][c]{%
    \includegraphics[width=0.23\textwidth]{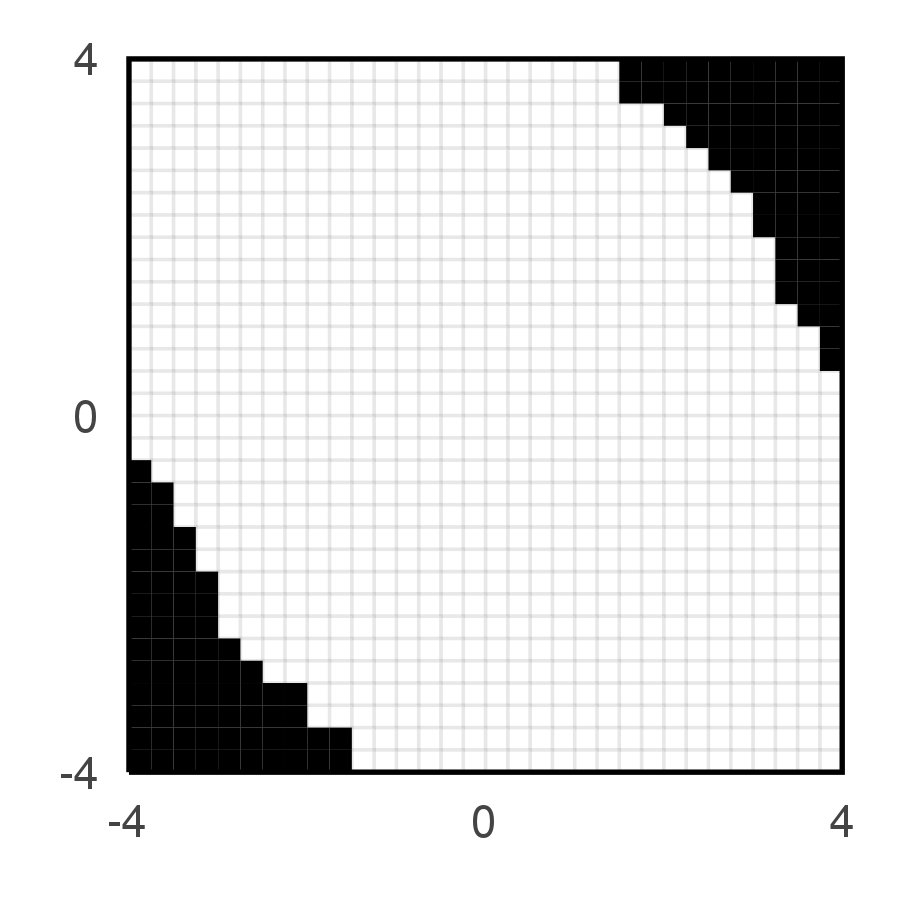}}
\caption{Minimum 2-step safety probability for $A_\text{rotation}$ for different values of $\distBound$.}
\label{fig:rotation-epsilon}
\end{figure}

The effect on the verification results of reducing $\distBound$ is shown in Figure~\ref{fig:rotation-epsilon} for $A_\text{rotation}$. The initial surely-safe areas diminish as $\distBound$ decreases until the minimum safety probability becomes zero nearly everywhere. This highlights a trade-off when choosing $\distBound$. Recall that Proposition~\ref{Th:Strategy} depends on enlarging and reducing the target set. Small $\distBound$ shrinks and enlarges the sets less, which can be beneficial when calculating the transition probabilities. Small $\distBound$ also tightens the bound on the distance between the system and the process. Too small, and the resulting probabilities become trivial everywhere, i.e., a minimum of zero and maximum of one. However, the choice of $\distBound$ is also upper-bounded by the size of the discretization of $\SafeSet$. 


\subsection{Switched Linear System}
\begin{figure}
    \centering
    \includegraphics[width=0.23\textwidth]{images/scale.png}\\
    \subcaptionbox{$T=1$ step}[0.23\textwidth][c]{%
    \includegraphics[width=0.23\textwidth]{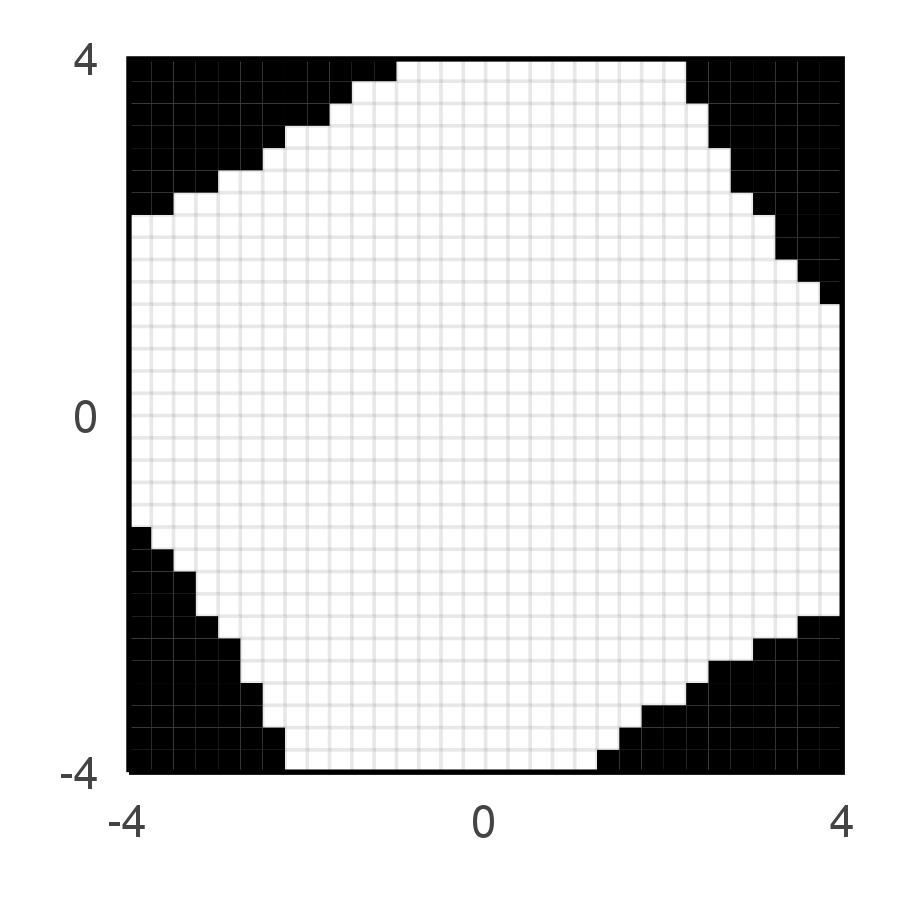}}
    \subcaptionbox{$T=1000$ steps}[0.23\textwidth][c]{%
    \includegraphics[width=0.23\textwidth]{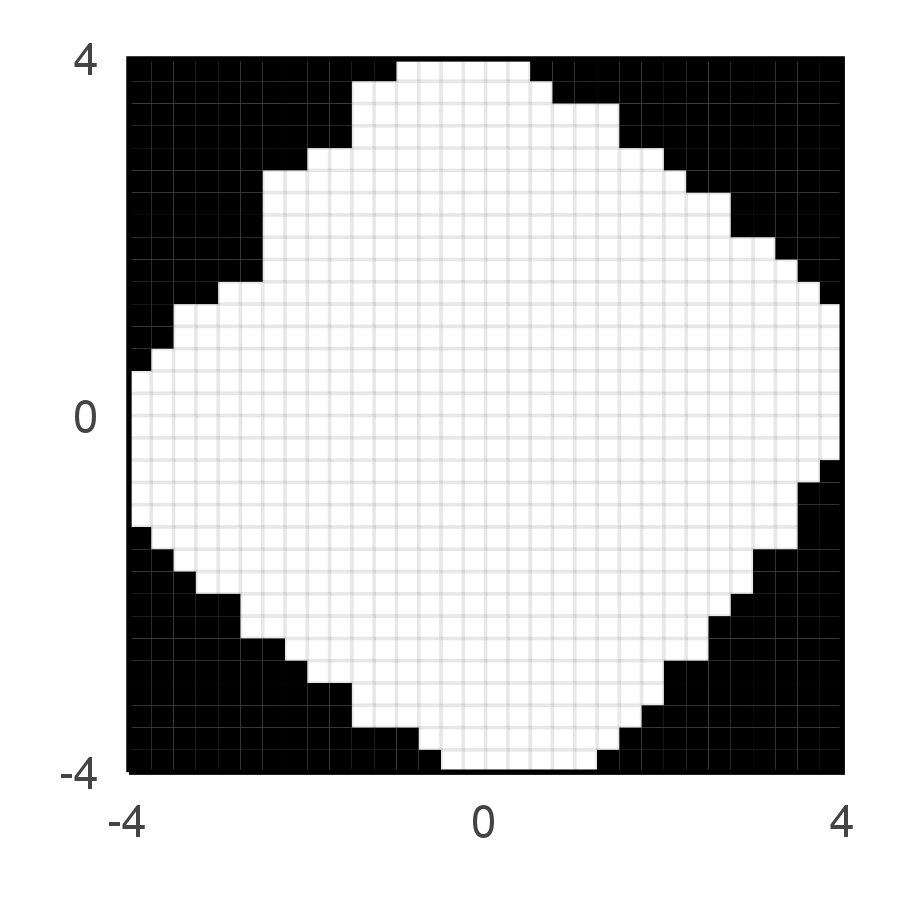}}
    \caption{Minimum probability of safety for a switched system comprised of the $A_\text{upper}$ and $A_\text{lower}$ systems.}
    \label{fig:linear-switched}
\end{figure}

The switched system uses the $A_\text{upper}$ and $A_\text{lower}$ systems from the previous section, and enables switching between the two at each time step.  The verification used the previous discretization of $\SafeSet$ and $\distBound=0.12$. 
Recall that the verification problem aims to check if this system remains in the safe set for all possible strategies. With two actions available to the system, the worst-case result occurs if one action could drive the system to an ``unsafe'' region of the other action.
Figure~\ref{fig:linear-switched} shows the minimum probability of safety after one and 1000 steps.
Due to the tight results of the component systems, the verification output of the switched system happens to be the superposition of the individual verification outputs. Notably, the system is guaranteed to remain in the safe set after 1000 steps \emph{regardless of the underlying strategy} so long it starts in a cell with a minimum safety probability of one.

\subsection{Nonlinear System}

\begin{figure}
    \qquad\subcaptionbox{True Vector Field}[0.25\textwidth][c]{%
    \includegraphics[width=0.25\textwidth]{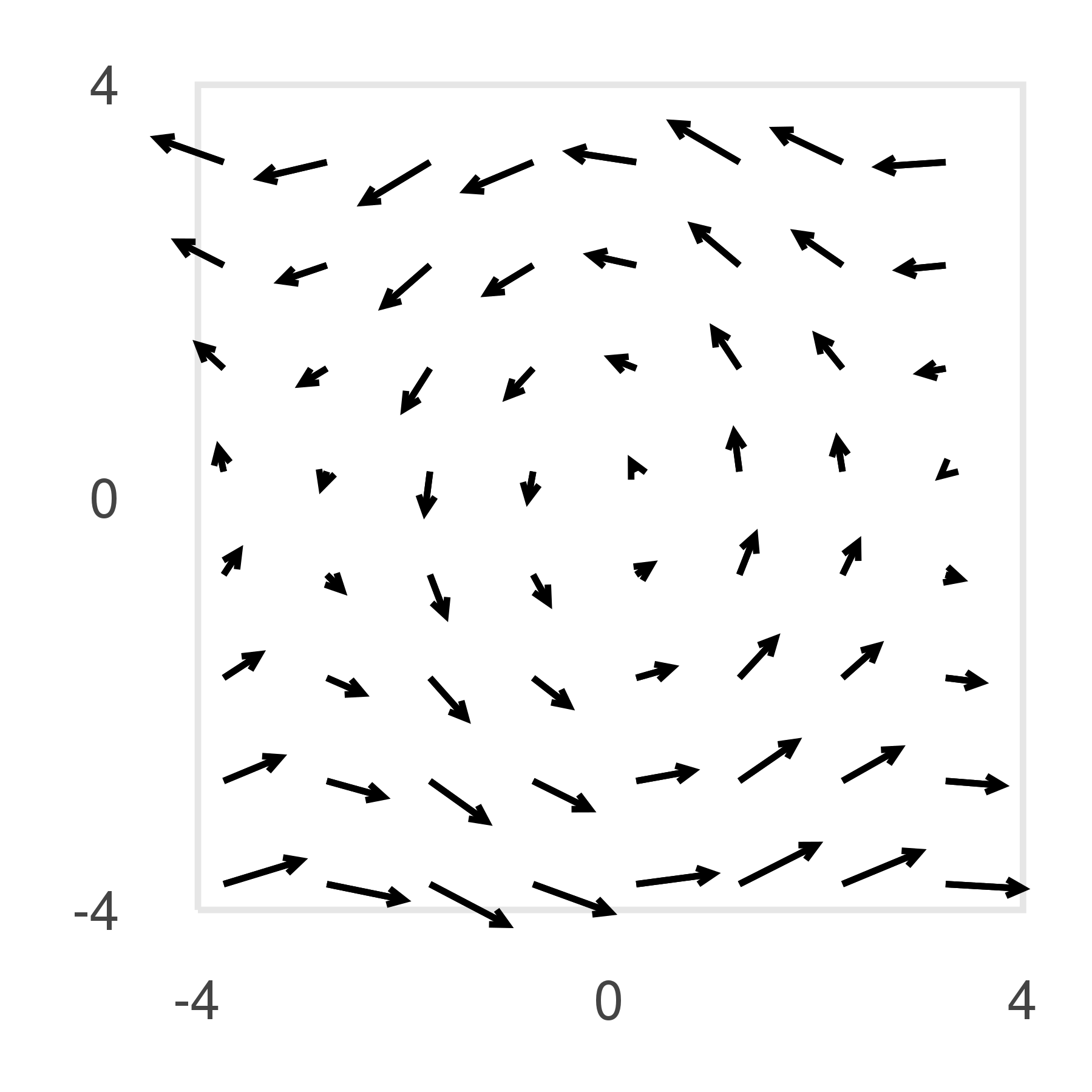}}\\
    \centering
    \includegraphics[width=0.23\textwidth]{images/scale.png}\\
    \subcaptionbox{$T=1$ step}[0.23\textwidth][c]{%
    \includegraphics[width=0.23\textwidth]{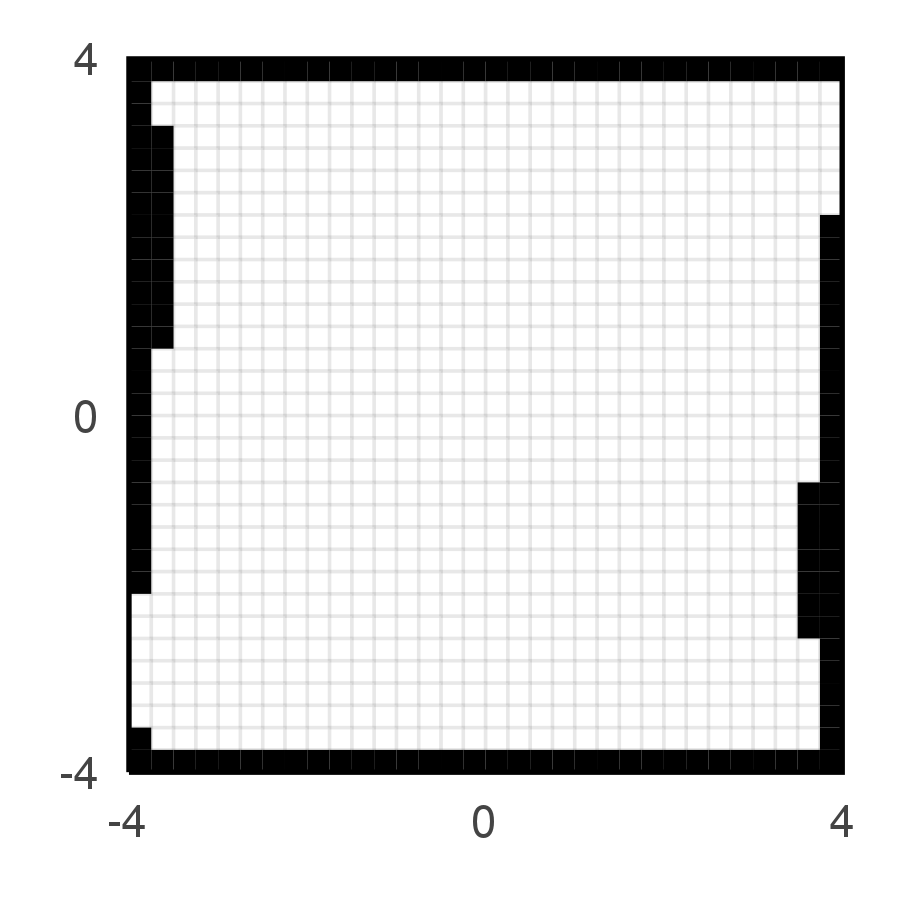}}
    \subcaptionbox{$T=2$ steps}[0.23\textwidth][c]{%
    \includegraphics[width=0.23\textwidth]{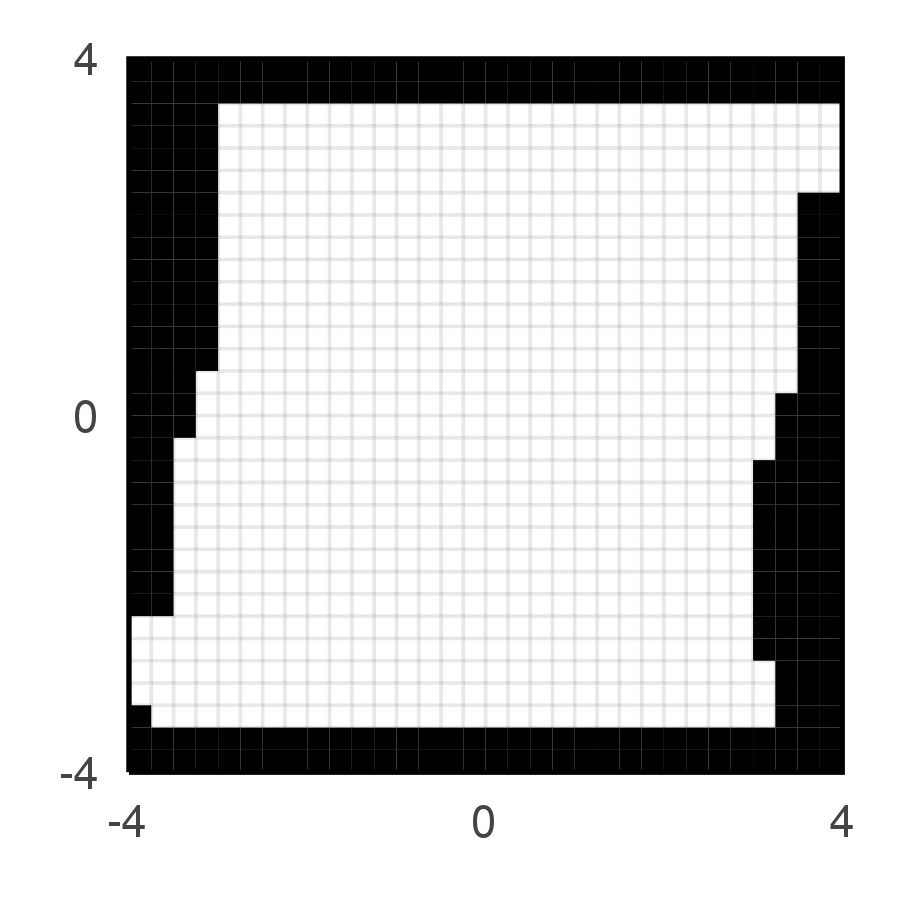}}\\\
    \subcaptionbox{$T=4$ steps}[0.23\textwidth][c]{%
    \includegraphics[width=0.23\textwidth]{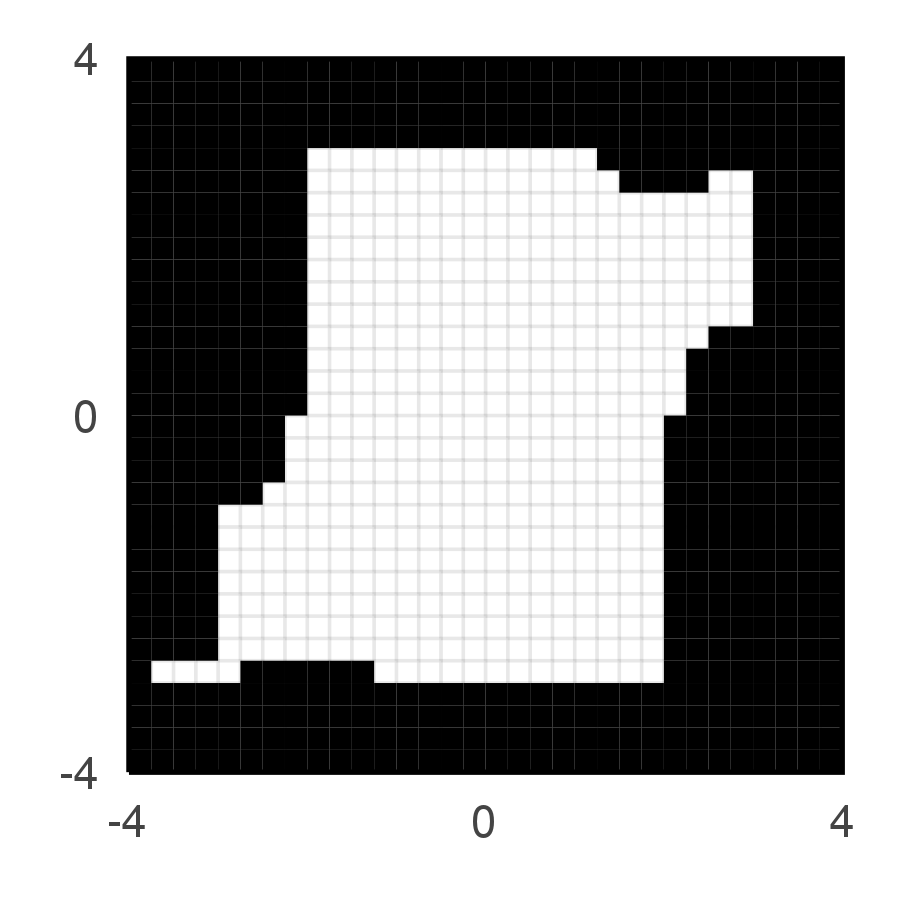}}
    \subcaptionbox{$T=6$ steps}[0.23\textwidth][c]{%
    \includegraphics[width=0.23\textwidth]{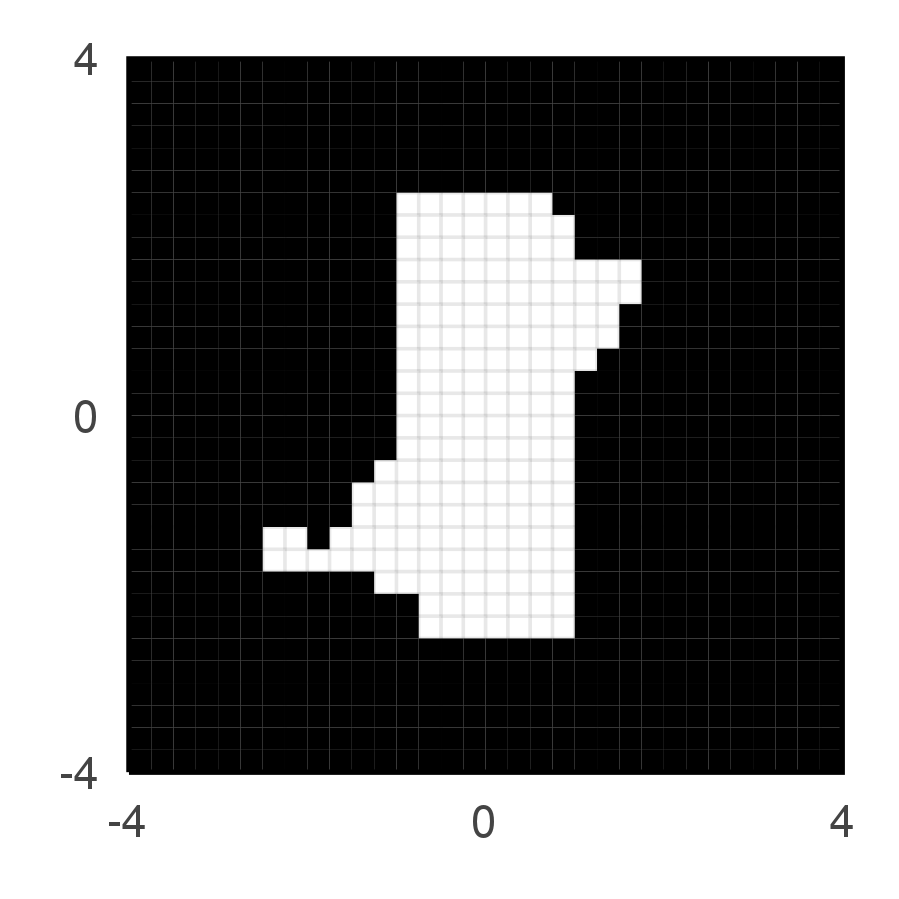}}
\caption{Vector fields and the minimum safety probability for multiple steps of the nonlinear system.}
    \label{fig:nonlinear}
\end{figure}

We demonstrate the verification on a nonlinear system given by
\begin{equation*}
    f(\x(k)) = [\,\x_1(k)-0.05\,\x_2(k),~ \x_2(k)+0.1\sin(\x_1(k))\,]^T
\end{equation*}
over a discretization of $\SafeSet$ with squares of side length 0.25. The vector field for the true system is shown in Figure~\ref{fig:nonlinear}(a). 
Many vectors flow away and out of $\SafeSet$ near parts of the border, while the field slowly spirals away from the origin. After 1 step, the minimum probability of safety is zero around parts of the field that flow out of $\SafeSet$ shown in Figure~\ref{fig:nonlinear}(b). However, the non-zero maximum probability of transitioning to parts of the field that flows out of $\SafeSet$ cause the initially-large set to shrink after successive steps. After 6 steps, safety can only be guaranteed if the system starts in regions around the origin. 



\section{CONCLUSION}
    \label{sec:conclusion}

We introduced a novel verification framework that generates safety guarantees for unknown dynamical systems. 
The approach is based on GP regression and an uncertain abstraction that incorporates probabilistic error bounds between the model learned from data and the underlying system.  
As a result, it allows the use of existing verification tools.  
This versatile framework paves the way for automatically generating guarantees for complex, safety-critical systems that have black-box components. 



\bibliographystyle{ieeetr}
\bibliography{cite,lahijanian}

\begin{thebibliography}{10}

\bibitem{Clarke99}
E.~M. Clarke, O.~Grumberg, and D.~Peled, {\em Model Checking}.
\newblock MIT Press, 1999.

\bibitem{BaierBook2008}
C.~Baier and J.-P. Katoen, {\em Principles of Model Checking}.
\newblock Cambridge, MA: The MIT Press, 2008.

\bibitem{tabuada2009verification}
P.~Tabuada, {\em Verification and control of hybrid systems: a symbolic
  approach}.
\newblock Springer Science \& Business Media, 2009.

\bibitem{Belta:Book:2017}
C.~Belta, B.~Yordanov, and E.~A. Gol, {\em Formal methods for discrete-time
  dynamical systems}, vol.~89.
\newblock Springer, 2017.

\bibitem{doyen2018verification}
L.~Doyen, G.~Frehse, G.~J. Pappas, and A.~Platzer, ``Verification of hybrid
  systems,'' in {\em Handbook of Model Checking}, pp.~1047--1110, Springer,
  2018.

\bibitem{kushner2013numerical}
H.~Kushner and P.~G. Dupuis, {\em Numerical methods for stochastic control
  problems in continuous time}, vol.~24.
\newblock Springer Science \& Business Media, 2013.

\bibitem{soudjani2015fau}
S.~E.~Z. Soudjani, C.~Gevaerts, and A.~Abate, ``Faust2: Formal abstractions of
  uncountable-state stochastic processes,'' in {\em International Conference on
  Tools and Algorithms for the Construction and Analysis of Systems},
  pp.~272--286, Springer, 2015.

\bibitem{Lahijanian:TAC:2015}
M.~Lahijanian, S.~B. Andersson, and C.~Belta, ``Formal verification and
  synthesis for discrete-time stochastic systems,'' {\em IEEE Transactions on
  Automatic Control}, vol.~60, pp.~2031--2045, Aug. 2015.

\bibitem{laurenti2020formal}
L.~Laurenti, M.~Lahijanian, A.~Abate, L.~Cardelli, and M.~Kwiatkowska, ``Formal
  and efficient synthesis for continuous-time linear stochastic hybrid
  processes,'' {\em IEEE Transactions on Automatic Control}, 2020.

\bibitem{Girard:ITAC:2007}
A.~Girard and G.~J. Pappas, ``Approximation metrics for discrete and continuous
  systems,'' {\em IEEE Transactions on Automatic Control}, vol.~52, no.~5,
  pp.~782--798, 2007.

\bibitem{Dutta:IFAC:2018}
S.~Dutta, S.~Jha, S.~Sankaranarayanan, and A.~Tiwari, ``Learning and
  verification of feedback control systems using feedforward neural networks,''
  {\em IFAC-PapersOnLine}, vol.~51, no.~16, pp.~151--156, 2018.

\bibitem{Haesaert:Automatica:2017}
S.~Haesaert, P.~M. Van~den Hof, and A.~Abate, ``Data-driven and model-based
  verification via bayesian identification and reachability analysis,'' {\em
  Automatica}, vol.~79, pp.~115--126, 2017.

\bibitem{Kenanian:Automatica:2019}
J.~Kenanian, A.~Balkan, R.~M. Jungers, and P.~Tabuada, ``Data driven stability
  analysis of black-box switched linear systems,'' {\em Automatica}, vol.~109,
  p.~108533, 2019.

\bibitem{Ahmadi:CDC:2017}
M.~Ahmadi, A.~Israel, and U.~Topcu, ``Safety assessemt based on
  physically-viable data-driven models,'' in {\em 2017 IEEE 56th Annual
  Conference on Decision and Control (CDC)}, pp.~6409--6414, IEEE, 2017.

\bibitem{rasmussen2003gaussian}
C.~E. Rasmussen, ``Gaussian processes in machine learning,'' in {\em Summer
  School on Machine Learning}, pp.~63--71, Springer, 2003.

\bibitem{cardelli2019robustness}
L.~Cardelli, M.~Kwiatkowska, L.~Laurenti, and A.~Patane, ``Robustness
  guarantees for bayesian inference with gaussian processes,'' in {\em
  Proceedings of the AAAI Conference on Artificial Intelligence}, vol.~33,
  pp.~7759--7768, 2019.

\bibitem{berkenkamp2015safe}
F.~Berkenkamp and A.~P. Schoellig, ``Safe and robust learning control with
  gaussian processes,'' in {\em 2015 European Control Conference (ECC)},
  pp.~2496--2501, IEEE, 2015.

\bibitem{srinivas2012information}
N.~Srinivas, A.~Krause, S.~M. Kakade, and M.~W. Seeger, ``Information-theoretic
  regret bounds for gaussian process optimization in the bandit setting,'' {\em
  IEEE Transactions on Information Theory}, vol.~58, no.~5, pp.~3250--3265,
  2012.

\bibitem{Germain2016pac}
P.~Germain, F.~Bach, A.~Lacoste, and S.~Lacoste-Julien, ``Pac-bayesian theory
  meets bayesian inference,'' in {\em Advances in Neural Information Processing
  Systems}, pp.~1884--1892, 2016.

\bibitem{chowdhury2017kernelized}
S.~R. Chowdhury and A.~Gopalan, ``On kernelized multi-armed bandits,'' in {\em
  Proceedings of the 34th International Conference on Machine Learning-Volume
  70}, pp.~844--853, JMLR. org, 2017.

\bibitem{lederer2019uniform}
A.~Lederer, J.~Umlauft, and S.~Hirche, ``Uniform error bounds for gaussian
  process regression with application to safe control,'' in {\em Advances in
  Neural Information Processing Systems}, pp.~657--667, 2019.

\bibitem{akametalu2014reachability}
A.~K. Akametalu, S.~Kaynama, J.~F. Fisac, M.~N. Zeilinger, J.~H. Gillula, and
  C.~J. Tomlin, ``Reachability-based safe learning with {G}aussian processes,''
  in {\em IEEE 53rd Annual Conference on Decision and Control (CDC), 2014:
  15-17 Dec. 2014, Los Angeles, California, USA}, pp.~1424--1431, IEEE, 2014.

\bibitem{sui2015safe}
Y.~Sui, A.~Gotovos, J.~W. Burdick, and A.~Krause, ``Safe exploration for
  optimization with gaussian processes,'' {\em Proceedings of Machine Learning
  Research}, vol.~37, pp.~997--1005, 2015.

\bibitem{berkenkamp2017safe}
F.~Berkenkamp, M.~Turchetta, A.~Schoellig, and A.~Krause, ``Safe model-based
  reinforcement learning with stability guarantees,'' in {\em Advances in
  neural information processing systems}, pp.~908--918, 2017.

\bibitem{polymenakos2019safety}
K.~Polymenakos, L.~Laurenti, A.~Patane, J.-P. Calliess, L.~Cardelli,
  M.~Kwiatkowska, A.~Abate, and S.~Roberts, ``Safety guarantees for planning
  based on iterative gaussian processes,'' {\em arXiv preprint
  arXiv:1912.00071}, 2019.

\bibitem{steinwart2001influence}
I.~Steinwart, ``On the influence of the kernel on the consistency of support
  vector machines,'' {\em Journal of machine learning research}, vol.~2,
  no.~Nov, pp.~67--93, 2001.

\bibitem{givan2000bounded}
R.~Givan, S.~Leach, and T.~Dean, ``Bounded-parameter {Markov} decision
  processes,'' {\em Artificial Intelligence}, vol.~122, no.~1-2, pp.~71--109,
  2000.

\bibitem{Hahn:QEST:2017}
E.~M. Hahn, V.~Hashemi, H.~Hermanns, M.~Lahijanian, and A.~Turrini,
  ``Multi-objective robust strategy synthesis for interval {M}arkov decision
  processes,'' in {\em Int. Conf. on Quantitative Evaluation of SysTems
  (QEST)}, (Berlin, Germany), pp.~207--223, Springer, Sep. 2017.

\bibitem{gaussianprocesses.jl}
J.~Fairbrother, C.~Nemeth, M.~Rischard, J.~Brea, and T.~Pinder,
  ``Gaussianprocesses. jl: A nonparametric bayes package for the julia
  language,'' {\em arXiv preprint arXiv:1812.09064}, 2018.

\end{thebibliography}


\end{document}